\documentclass{article}

\usepackage{PRIMEarxiv}

\usepackage[utf8]{inputenc} 
\usepackage[T1]{fontenc}    
\usepackage{hyperref}       
\usepackage{url}            
\usepackage{booktabs}       
\usepackage{amsfonts}       
\usepackage{nicefrac}       
\usepackage{microtype}      
\usepackage{lipsum}
\usepackage{amssymb}
\usepackage{fancyhdr}       
\usepackage{graphicx}       
\graphicspath{{media/}}     

\usepackage{multirow} 
\usepackage{array} 

\usepackage{amsmath} 
\usepackage{xcolor}

\usepackage{natbib}
\usepackage{amsthm}

\newtheorem{theorem}{Theorem}
\newtheorem{proposition}{Proposition}

\usepackage{caption}
\usepackage{subcaption}

\usepackage{algorithm}
\usepackage{algorithmic}

\title{MDAS: A Diagnostic Approach to Assess the Quality of Data Splitting in Machine Learning
}

\author{
  Palash Ghosh \\
    \\
Department of Mathematics \\
  Indian Institute of Technology Guwahati, India\\
  and \\
  Centre for Biomedical Data Science\\
  Duke-NUS Medical School\\
  National University of Singapore, Singapore \\
  \\
  \And
  Bittu Karmakar \\
  \\
  Department of Mathematics \\
  Indian Institute of Technology Guwahati, India \\
  \\
  \And
  Eklavya Jain \\
  \\
  Department of Mathematics \\
  Indian Institute of Technology Guwahati, India \\
  \\
  \And
  J. Neeraja \\
  \\
  Department of Mathematics \\
  Indian Institute of Technology Guwahati, India \\
  \\
  \And
  Buddhananda Banerjee \\
  \\
  Department of Mathematics and Center for Excellence in AI\\
  Indian Institute of Technology Kharagpur, West Bengal, India\\
  \\
  \And
  Tanujit Chakraborty \\
    \\
  SAFIR \\
  Sorbonne University Abu Dhabi, United Arab Emirates\\
  and \\
  Sorbonne Centre for Artificial Intelligence\\
  Sorbonne University, Paris, France\\
  \\
  Correspondence to: tanujit.chakraborty@sorbonne.ae\\
}

\begin{document}

\maketitle

\begin{abstract}
In the field of machine learning, model performance is usually assessed by randomly splitting data into training and test sets. Different random splits, however, can yield markedly different performance estimates, so a genuinely good model may be discarded or a poor one selected purely due to an unlucky partition. This motivates a principled way to diagnose the quality of a given data split. We propose a diagnostic framework based on a new discrepancy measure, the Mahalanobis Distribution Alignment Score (MDAS). MDAS is a symmetric dissimilarity measure between two multivariate samples, rather than a strict metric. MDAS captures both mean and covariance differences and is affine invariant. Building on this, we construct a Monte Carlo test that evaluates whether an observed split is statistically compatible with typical random splits, yielding an interpretable p-value for split quality. Using several real data sets, we study the relationship between MDAS and model robustness, including its association with the normalized Akaike information criterion. Finally, we apply MDAS to compare existing state-of-the-art deterministic data-splitting strategies with standard random splitting. The experimental results show that MDAS provides a simple, model-agnostic tool for auditing data splits and improving the reliability of empirical model evaluation.
\end{abstract}


\section{Introduction} \label{sec:intro}
Statistical and machine learning models are widely used for inference and prediction tasks. The goal of inference is to understand or test hypotheses about how a system behaves, whereas prediction aims to forecast unobserved outcomes or future behavior \cite{bzdok2018krzywinski, zellner2004statistics}. When the sole objective is to infer associations (or causality), data are typically not split into training and test sets. For example, in a randomized controlled trial (RCT), researchers use the entire dataset to determine whether a treatment is more effective than an alternative treatment or placebo \cite{friedman2015fundamentals, liu2025thompson}. In contrast, in supervised machine learning tasks, where the goal is prediction, it is standard practice to split the data into training and test sets for model development and evaluation, respectively \cite{stone1974cross, hastie2009elements}. The model is first fitted on the training data by estimating its parameters or functions, and the resulting model is then evaluated on the test data. The resulting test performance is taken as a proxy for how well the model will generalize to future and unseen data. Various model selection and hyperparameter tuning procedures are built entirely around this principle. Thus, the way in which the data are split into training and test sets plays a central role in practical machine learning workflows, yet the split itself is often treated as a routine, almost invisible step \cite{picard1990data, reitermanova2010data}. This tension between the importance of the split and the ad hoc way in which it is usually chosen motivates a closer examination of what constitutes a ``good'' data split and how its quality can be assessed quantitatively.

Birba \emph{et al.} present a comprehensive comparison of various data splitting methods employed in machine learning and demonstrate how different splitting strategies affect the estimation of a model’s generalizing ability \cite{birba2020comparative}. They experiment with techniques such as k-fold cross-validation, bootstrap-based random splitting, the Kennard–Stone (K-S) algorithm, and the SPXY algorithm \cite{kohavi1995study, kennard1969computer, GALVAO2005736}. They conclude that data splitting remains a heuristic step and that its relationship with model performance is strongly dependent on the underlying data \cite{birba2020comparative}. A widely accepted notion for obtaining reliable performance estimates is that the training and test sets should adequately represent the entire dataset \cite{sadhukhan2024footprints}. The K-S algorithm and its successor SPXY are built on this notion and aim to preserve the original data structure in the selected subsets \cite{kennard1969computer, GALVAO2005736}. Both CADEX (K-S) and SPXY rely on an underlying distance metric, typically the Euclidean distance. The key difference is that SPXY considers the statistical variation of the dependent variable along with the independent variables when selecting representative subsets, whereas CADEX only considers the independent variables \cite{GALVAO2005736}. Galvão \emph{et al.} argue that this inclusion leads to a more effective distribution of samples in the multidimensional space and thereby enhances predictive performance \cite{GALVAO2005736}. Joseph and Vakayil propose a new data splitting method (SPlit) based on support points and compare it with the deterministic CADEX (K-S) and DUPLEX algorithms \cite{joseph2021split}. The SPlit method follows a similar idea when partitioning data into training and test sets: it first identifies the most representative points for testing and uses the remaining samples for training \cite{joseph2021split}. They compute optimal representative points, or Support Points (SP), for the entire dataset and then employ a nearest-neighbors strategy to sequentially subsample, reporting substantial improvements in worst-case predictions compared with CADEX and DUPLEX \cite{joseph2021split}. However, choosing a test set that closely mimics the entire dataset is not necessarily a good policy for concluding model robustness, since even a poor model can perform well on a carefully engineered split. Xu \emph{et al.} find that K-S and SPXY can yield poor estimates of model performance, because the most representative samples are chosen first, leaving a poorly representative subset for performance evaluation \cite{xu2018splitting}.

Existing splitting methods aim to divide a dataset into training and test sets that share the same distribution as the original data \cite{kahloot2021algorithmic, joseph2021split, reitermanova2010data, babaei2025impact}. We call such a partition an \textit{ideal} data split. But do we actually need that kind of split? In practice or production, model performance is evaluated on new data that will typically differ, at least to some extent, from the data used during development \cite{varoquaux2023evaluating, vamathevan2019applications}. Ideally, one would like to test a model using both a dataset whose distribution closely matches that of the training data and another dataset whose distribution deviates (reasonably) from it \cite{altalhan2025imbalanced}. Good performance on the former type of data indicates that the model behaves well when future data are drawn from essentially the same distribution as the training data; poor performance suggests that the model is unsuitable for deployment and may be discarded \cite{zha2025data}. In contrast, we do not expect equally strong performance in the second testing environment, where the distribution has shifted. If the model still achieves reasonable performance in this setting, it provides evidence that the model is robust to perturbations in the distribution of new data \cite{li2025machine}.
 
While much of the existing work has focused on constructing an ideal data split for model building and evaluation, the quality of a particular realized split has not been extensively analyzed, partly because most data splitting methods are heuristic in nature. In this light, this paper offers a new perspective on data splitting. We propose a diagnostic approach built around a new distance-based test statistic, the Mahalanobis Distribution Alignment Score (MDAS), which is based on the Mahalanobis squared distance, followed by a hypothesis test to assess the quality of a data split, whether random or non-random \cite{mclachlan1999mahalanobis}. The MDAS statistic, denoted by $\Lambda$, quantifies the multivariate distance between the training and test sets. A key advantage of this approach is that it assesses the quality of a split without requiring any specification of the predictive model to be fitted later. The accompanying Monte Carlo simulation-based hypothesis test evaluates whether the training and test sets can be regarded as coming from similar distributions. In addition, when a specific model (for example, a regression model) has been chosen, we also illustrate, via graphical summaries, the relative performance of that model on the given split compared with all possible splits, using the normalized Akaike Information Criterion (AIC) \cite{sakamoto1986akaike}.

We outline the main contributions of this paper as follows:
\begin{enumerate}
    \item We introduce the MDAS: Mahalanobis Distribution Alignment Score, a symmetrized Mahalanobis-based distance between training and test sets, and study its mathematical properties, including non-negativity, symmetry, affine invariance, decomposition into mean- and covariance-mismatch components, and consistency properties.
    \item We develop a Monte Carlo hypothesis testing framework based on MDAS to quantitatively assess the quality of any given train–test split (random or deterministic), providing an interpretable p-value for split quality and establishing its asymptotic behaviour under the null and alternative.
    \item We propose a graphical diagnostic tool that links MDAS to model performance, using the normalized AIC to position the observed split relative to all possible splits for a chosen model, thus connecting distributional alignment to model robustness.
    \item We conduct an empirical study on real datasets, comparing random splitting with deterministic strategies such as CADEX (K-S), DUPLEX, and SPlit, and demonstrate how MDAS can be used to audit and compare data-splitting methods in terms of both distributional alignment and predictive performance.
\end{enumerate}

The paper is organized as follows. Section \ref{sec:Different-Splittings} describes different data splitting strategies with illustrative examples. Section \ref{sec:methodology} presents the proposed methodology, formulates the hypotheses, and details the algorithm. The results of various experiments on real datasets, together with a discussion of critical observations, are reported in Section \ref{sec:experiments}. Finally, Section \ref{sec:conclusions} summarizes the main findings and comments on the applicability and limitations of the proposed approach.

\section{Splittings Strategies and Motivating Examples}
\label{sec:Different-Splittings}
\begin{figure*}[ht]
    \centering
    \includegraphics[width = 0.90\linewidth]{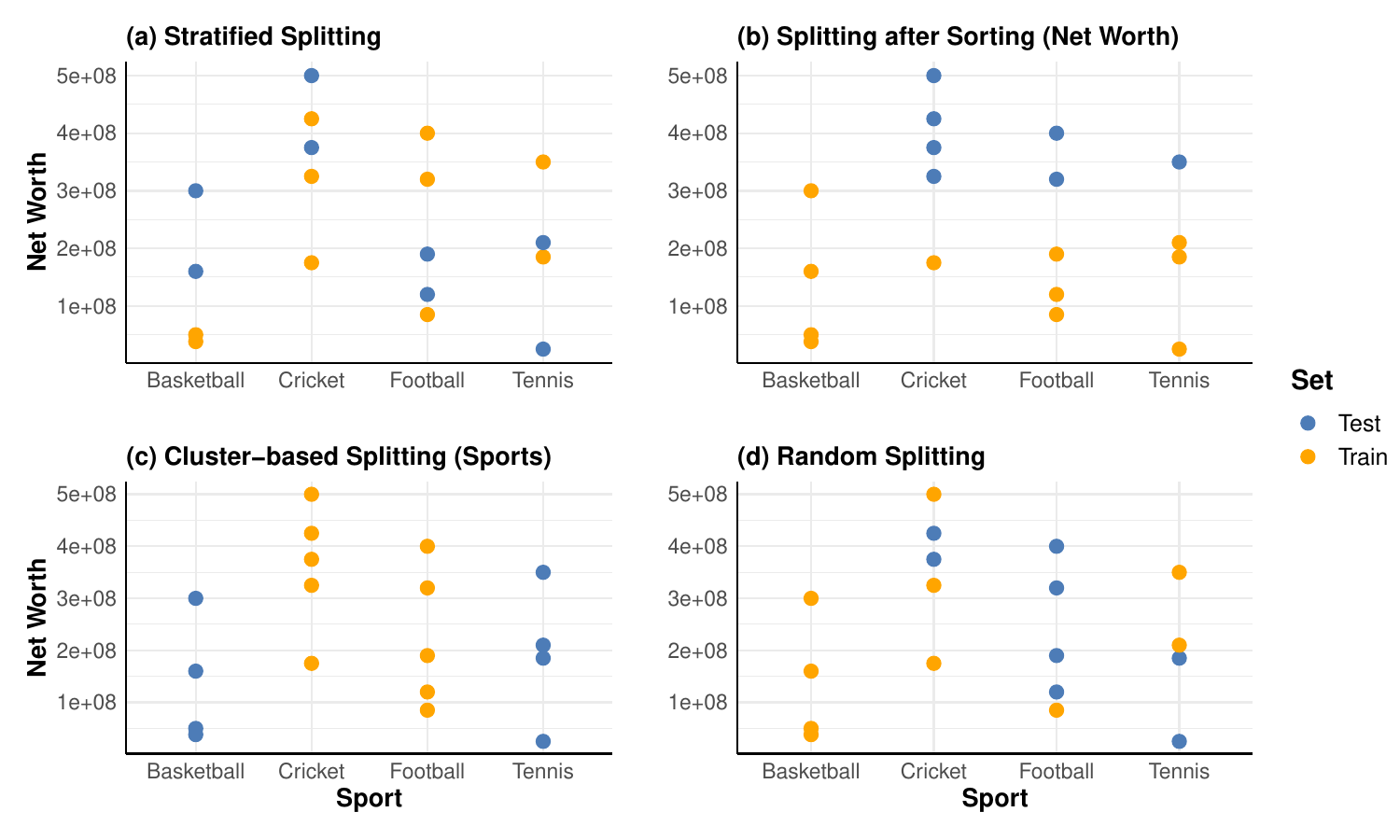}
    \caption{\textbf{Data splitting strategies.} Each subplot represents a splitting technique. Each ball in a subplot corresponds to a player. {\color{blue} Blue} (dark) / {\color{orange} orange} (bright) balls represent examples for {\color{blue} test} / {\color{orange} training}. The split percentage is considered to be 50\% for cluster-based splitting, and 60\% for the remaining.}
    \label{fig:splitting}
\end{figure*}

The simplest and perhaps most common strategy to split a dataset and obtain the corresponding training set (and test set) is to sample a fraction (say 80\%) of the dataset randomly \cite{l1991implementing}. This strategy is referred to as random splitting, and it sometimes leads to a heavily fragmented decision boundary \cite{ishwaran2015effect}. Other techniques like cluster-based splitting, stratified splitting, and adversarial or biased splitting \cite{sogaard2020we} can be used and are examined in Fig. \ref{fig:splitting}. The underlying data is a hypothetical dataset of sports players and their net worth. We assume an objective of associating a player's net worth with the sport they play and compare the four data splitting strategies by plotting Sport vs. Net Worth. It is expected that different splitting techniques will produce different train-test partitions for a given split percentage. Stratified splitting or stratified random sampling obtains a sample population that best represents the entire population under investigation. Consequently, in Fig. \ref{fig:splitting}a, approximately 60\% entries from each sport are randomly chosen for training while the remaining are kept for testing purposes. Class imbalance achieved after partitioning the data sorted on net worth, as shown in Fig. \ref{fig:splitting}b, is a classic example of introducing an adversarial effect. Adversarial splits are a great way to examine the true capability of a model. Søgaard et al. \cite{sogaard2020we} conclude that multiple biased splits give a more realistic estimation of out-of-sample error as compared to multiple random splits. Another typical technique to split a dataset is cluster-based splitting. In Fig. \ref{fig:splitting}c, the complete dataset is split by forming clusters of sports. Considering a split percentage of 50\%, we randomly assign 2 clusters to the training set and the remaining 2 clusters to the test set. Finally, Fig. \ref{fig:splitting}d portrays random splitting where no restrictions are in place. All data points are pooled together and split into two subsets comprising 60\% (training) and 40\% (test) of the data, respectively. In particular, a random split with 60\% split percentage can result in the following outcomes:
\begin{enumerate}
    \item {\textit{More than 60\% entries from a sport in the training set (Basketball),}}
    \item {\textit{Less than 60\% entries from a sport in the training set (Football),}}
    \item {\textit{Exactly 60\% entries from a sport in the training set (Cricket).}}
\end{enumerate}
 
\begin{figure*}[t]
    \centering
    \tiny
    \begin{subfigure}[b]{0.35\textwidth}
        \includegraphics[width=\linewidth]{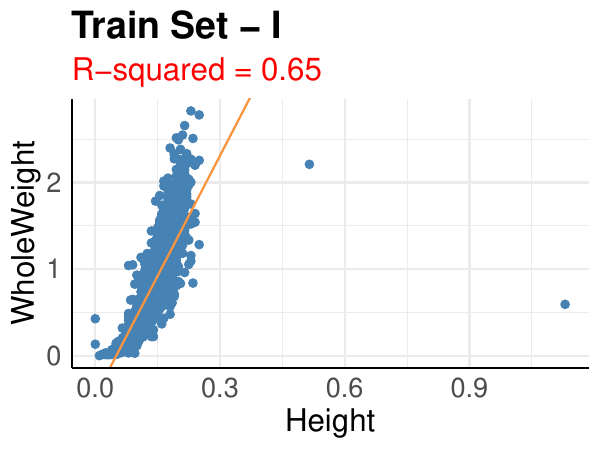}
        \captionsetup{font=footnotesize}
        \caption{\textbf{First Simulation [Train].} Both outliers are present in the train set.}\label{fig:train normal1}
    \end{subfigure}
    \hfill
    \begin{subfigure}[b]{0.35\textwidth}
        \includegraphics[width=\linewidth]{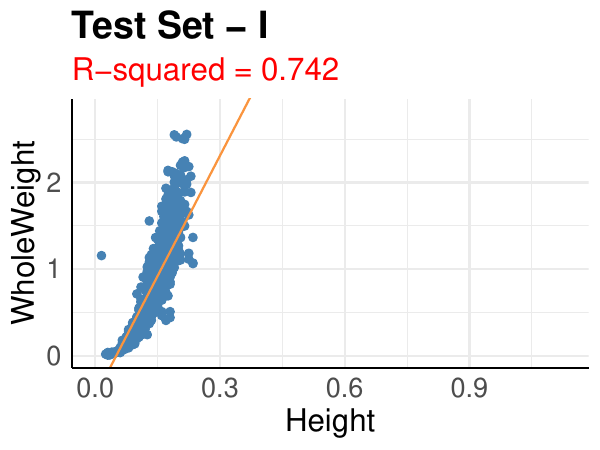}
        \captionsetup{font=footnotesize}
        \caption{\textbf{First Simulation [Test].} A higher $R^2$ value for the test set implies good model performance.}
        \label{fig:test normal1}
    \end{subfigure}
    \vspace{0.5em}

    \begin{subfigure}[b]{0.35\textwidth}
        \includegraphics[width=\linewidth]{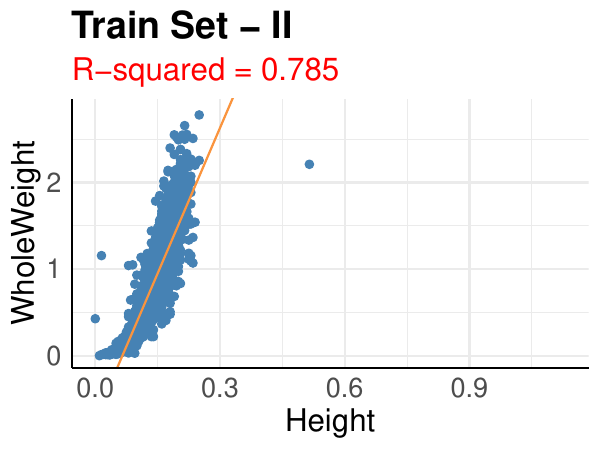}
        \captionsetup{font=footnotesize}
        \caption{\textbf{Second Simulation [Train].} Only one outlier is present in the train set.}\label{fig:train drop1}
    \end{subfigure}
    \hfill
    \begin{subfigure}[b]{0.35\textwidth}
        \includegraphics[width=\linewidth]{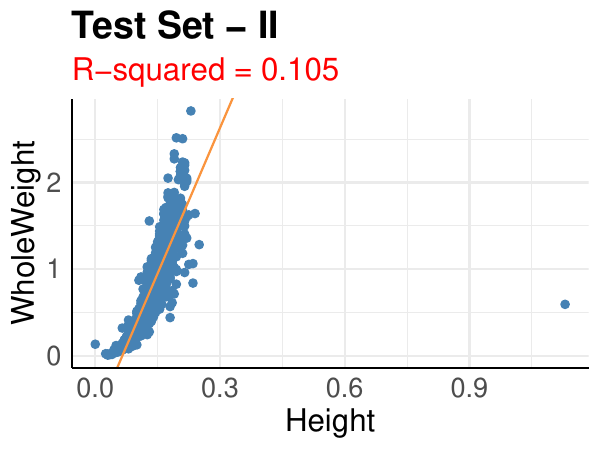}
        \captionsetup{font=footnotesize}
        \caption{\textbf{Second Simulation [Test].} Much lower $R^2$ value for the test set, implying poor model performance.}
        \label{fig:test drop1}
    \end{subfigure}

    \caption{\textbf{Drop in model performance:} The objective is to regress the weight of the species with their height. The data is represented using {\color{blue} blue} dots. The {\color{orange} orange} line represents the linear regression model fitted on train set.}\label{fig:drop in performance1}
\end{figure*}

Fig. \ref{fig:splitting} shows how random splitting differs from other techniques and highlights its indifferent behavior towards maintaining similar distributions of the training and test sets. The randomness induced by the random splitting method generally eliminates subtle biases that impede a conclusive evaluation of the model. Ideally, the random split should maintain the same statistical distribution of the original data in the training and test data. However, in practice, we may see the distributions in the training and test vary a lot after random splitting. Fig. \ref{fig:drop in performance1} presents two scenarios of random splitting of the original data. To illustrate how random splitting can highly influence the model assessment, we fit a regression model into the training data and assess its performance in the test data. We consider coefficient of determination ($R^2$) values to assess the model performance. Fig. \ref{fig:drop in performance1} uses the Abalone Dataset from the UCI Machine Learning Repository to regress the weights of 4177 abalone fishes with their heights. In the first scenario, the first row of Fig. \ref{fig:drop in performance1}, the test $R^2$ is much higher than the training $R^2$, indicating a good model fit. In the second row, the test $R^2$ drops compared to the corresponding training $R^2$. The drop in the model performance is a consequence of the position of two apparent outliers \cite{osborne2004power}, without going into further detail to decide whether they are influential or outlier points. In summary, we observe that when both the outliers are in the training set (Fig. \ref{fig:train normal1}), the corresponding test set reports a higher $R^2$ (Fig. \ref{fig:test normal1}). In contrast, the presence of an outlier in the test set (Fig. \ref{fig:test drop1}) and the other in the training set results in a drop in the $R^2$ value. This example shows that two simulations yield significantly different model performances for the same model relation. Since the behavior is ambiguous and depends on the underlying data split that led to the variation, it is difficult to estimate the correct model performance.

\begin{figure*}[t]
    \centering
    \tiny
    \begin{subfigure}[b]{0.35\textwidth}
        \includegraphics[width=\linewidth]{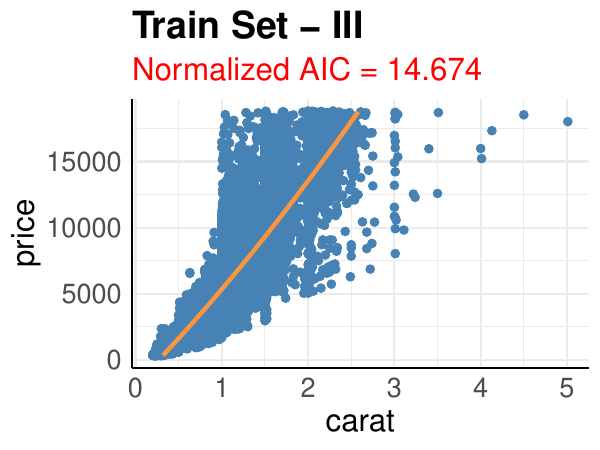}
        \captionsetup{font=footnotesize}
        \caption{\textbf{First Simulation [Train].}}\label{fig:train normal2}
    \end{subfigure}
    \hfill
    \begin{subfigure}[b]{0.35\textwidth}
        \includegraphics[width=\linewidth]{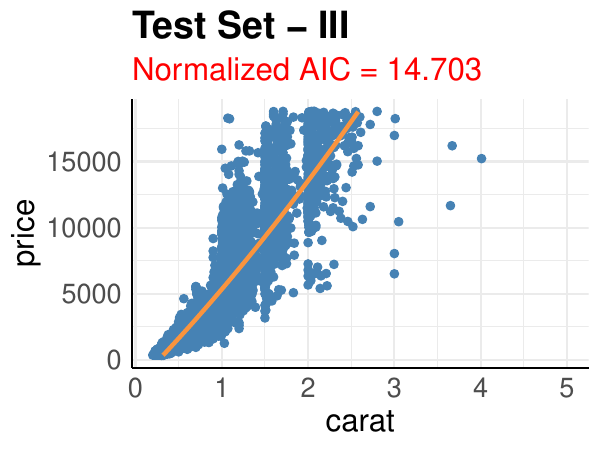}
        \captionsetup{font=footnotesize}
        \caption{\textbf{First Simulation [Test].} Similar AIC score for the test set as compared to the train set, implying a good model performance.}\label{fig:test normal2}
    \end{subfigure}
    \vspace{0.5em}

    \begin{subfigure}[b]{0.35\textwidth}
        \includegraphics[width=\linewidth]{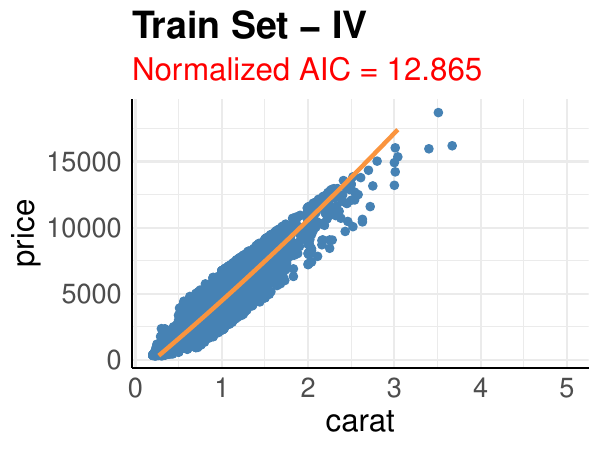}
        \captionsetup{font=footnotesize}
        \caption{\textbf{Second Simulation [Train].} }\label{fig:train drop2}
    \end{subfigure}
    \hfill
    \begin{subfigure}[b]{0.35\textwidth}
        \includegraphics[width=\linewidth]{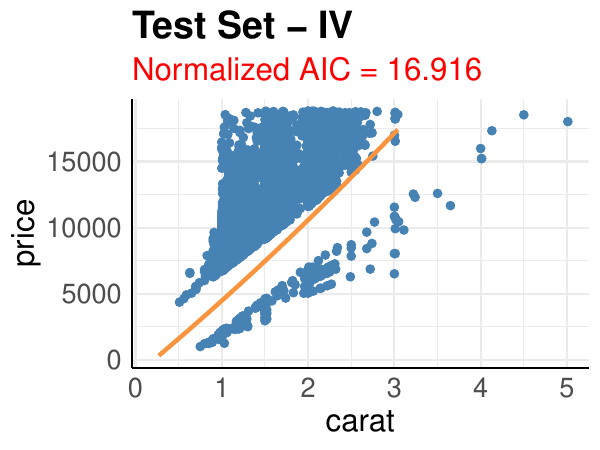}
        \captionsetup{font=footnotesize}
        \caption{\textbf{Second Simulation [Test].} Much higher AIC score for the test set as compared to the train set, implying a poor model performance.}
        \label{fig:test drop2}
    \end{subfigure}

    \caption{\textbf{Drop in model performance:} The objective is to associate the price of the diamonds with the carats of the diamond. The data is represented using {\color{blue} blue} dots. The {\color{orange} orange} curve represents the polynomial regression model fitted on train set. }
        \label{fig:drop in performance2}
\end{figure*}

Another example, shown in Fig. \ref{fig:drop in performance2}, portrays a similar ambiguous behavior in estimating model performance. The dataset used in Fig. \ref{fig:drop in performance2} is the Diamonds dataset with 53,940 entries available in the ggplot library in R. We attempt to map out the price of the diamond based on the carat of the diamond used. We use a polynomial regression model and the normalized AIC score as discussed in Section \ref{subsec:naic}. A drop (higher AIC) in the test model performance signifies that random splitting can lead to ambiguous model performances. Consider the first simulation for this dataset in Fig. \ref{fig:train normal2} where the split percentage is 80\%. We observe strong model performance on both the test and training splits when comparing two models; the one with the lower normalized AIC score is considered more robust. Although in the second simulation, the normalized AIC score goes to 16.916 due to an adversarial data split, indicating a poor model performance on the test split when compared with the corresponding train split, as well as the previous simulation. 

Therefore, different random splits can lead to different model performances, making any conclusions about model robustness unreliable. To tackle this selection bias or prevent overfitting, researchers have employed other model evaluation techniques like cross-validation \cite{refaeilzadeh2009cross}, which provide efficient estimates by averaging the model performance over a number of train-test splits. However, this method forces us to fit the model on different training datasets repeatedly. 

Thus, it raises the need for a method to correctly estimate the model performance without re-training the model or even knowing the model relation. Since the variation in model performance is primarily precipitated by the random splitting step, we propose a statistical technique to diagnose and classify splits as ``good" or ``bad". A ``good'' split would yield reliable model performances, while a bad split would not. Before we proceed to the method, we discuss the Mahalanobis squared distance in the next section.

\section{Methodology}
\label{sec:methodology}

\subsection{Background: Mahalanobis Squared Distance}
\label{sec:Mahalanobis squared distance}

Distance measures are essential components of numerous machine learning techniques \cite{xiang2008learning}. Learning algorithms like KMeans \cite{krishna1999genetic} and K-nearest neighbor (KNN) \cite{peterson2009k} are supported by such metrics due to their need for a suitable distance metric for identifying neighboring points. One such widely used distance measure is the Mahalanobis squared distance \cite{mclachlan1999mahalanobis}. The Mahalanobis squared distance is the distance of an observation $\boldsymbol{x}$ from a set of observations with mean vector $\boldsymbol{\mu}$, and a non-singular pooled covariance matrix $\Sigma$. It is expressed as
\begin{equation}
\label{eqn:Mahalanobis squared distance}
    \Delta^2 = (\boldsymbol{x} - \boldsymbol{\mu})^T\Sigma^{-1}(\boldsymbol{x} - \boldsymbol{\mu}).
\end{equation}
The use of Mahalanobis squared distance has grown over the years. It is used in data clustering \cite{xiang2008learning}, image segmentation \cite{zhang2011image}, incremental learning \cite{yu2025generating}, and face pose estimation problems. A modified version of the distance is used in classification tasks executed through K-nearest neighbors in a multivariate setup \cite{doi:10.1080/00401706.2014.902774}. The Mahalanobis squared distance takes into account correlations and scales of variables \cite{brereton2016re} and it is also used for outlier detection \cite{geun2000multivariate}.

The basic notion for obtaining a good train-test split is to make the distribution of the training set and the test set close to each other. In other words, the farther the test distribution from the training distribution, the less accurate the estimate of model generalizability is reported. Building on this notion, we try to find the distance between the training and the test sets. Inspired by the pervasive use of the Mahalanobis squared distance, initially proposed by Mahalanobis \cite{mahalanobis1936generalized}, in statistical as well as machine learning tasks, we define a new distance measure based on it. We use it to quantify the distance between any two population samples, more particularly, training and test samples. We use this distance measure to ultimately diagnose the quality of a random split irrespective of the model relation and the problem type through Monte Carlo simulation-based hypothesis testing \cite{theiler1996constrained} discussed in the next section.

\subsection{Mahalanobis Distribution Alignment Score (MDAS)}
As discussed in Section \ref{sec:Mahalanobis squared distance}, we use the Mahalanobis squared distance to calculate the distance between two multivariate populations. Consider $\boldsymbol{X} = (\boldsymbol{x_1}, \boldsymbol{x_2}, \dots, \boldsymbol{x_{n_X}})^T$ and $\boldsymbol{Y} = (\boldsymbol{y_1}, \boldsymbol{y_2}, \dots, \boldsymbol{y_{n_Y}})^T$ to be the two data samples with $n_X$ and $n_Y$ be the number of observations, respectively. Let $\boldsymbol{\mu_X}$ and $\boldsymbol{\mu_Y}$ be the means of the two corresponding populations. Each observation $\boldsymbol{x}_i$ or $\boldsymbol{y}_j$ is a $p$-dimensional feature vector in $\mathbb{R}^p$, and hence all covariance matrices (e.g., $\Sigma_X$, $\Sigma_Y$, $\Sigma_p$) are of dimension $p \times p$, where $p$ denotes the number of variables (features). Further, obtain the pooled variance-covariance matrix \cite{seber2009multivariate} of entire data as,
\begin{equation}
\label{eqn:pooled covariance}
    \Sigma_p = \frac{(n_X - 1)\Sigma_X + (n_Y - 1)\Sigma_Y}{n_X + n_Y - 2}
\end{equation}
From (\ref{eqn:Mahalanobis squared distance}), we calculate the distance of each observation $x_i$ from the other population $\boldsymbol{Y}$ using (\ref{eqn:d1}). Similarly, we calculate the distance of each observation $y_j$ from the population $\boldsymbol{X}$ using (\ref{eqn:d2}). The two distances are given as 
\begin{equation}
    \label{eqn:d1}
    \Delta_{x_iY}^2 =(\boldsymbol{x_i} - \boldsymbol{\mu_Y})^T\Sigma_p^{-1}(\boldsymbol{x_i} - \boldsymbol{\mu_Y}),\ \forall i = 1 \dots n_X,
\end{equation}
\begin{equation}
    \label{eqn:d2}
    \Delta_{y_jX}^2 =(\boldsymbol{y_j} - \boldsymbol{\mu_X})^T\Sigma_p^{-1}(\boldsymbol{y_j} - \boldsymbol{\mu_X}),\ \forall j = 1 \dots n_Y,
\end{equation}
respectively. We assume that the training and the test data sets have the same variance-covariance structure and it can be represented by (\ref{eqn:pooled covariance}). The two expressions in (\ref{eqn:d1}) and (\ref{eqn:d2}) calculate the distance of a single observation of $\boldsymbol{X}$ from $\boldsymbol{Y}$ and of $\boldsymbol{Y}$ from $\boldsymbol{X}$, respectively. Further, we define the average distance of population $\boldsymbol{X}$ from population $\boldsymbol{Y}$, and vice-versa as,
\begin{equation}
    \label{eqn:avg d1}
    \Delta_{XY}^2 = \frac{1}{n_X} \sum_{i = 1}^{n_X}(\boldsymbol{x_i} - \boldsymbol{\mu_Y})^T\Sigma_p^{-1}(\boldsymbol{x_i} - \boldsymbol{\mu_Y}),
\end{equation}
\begin{equation}
    \label{eqn:avg d2}
    \Delta_{YX}^2 = \frac{1}{n_Y} \sum_{j = 1}^{n_Y} (\boldsymbol{y_j} - \boldsymbol{\mu_X})^T\Sigma_p^{-1}(\boldsymbol{y_j} - \boldsymbol{\mu_X}),
\end{equation}
respectively. Finally, we calculate our distance, called MDAS ($\Lambda$), as the average of the newly defined distances, $\Delta_{XY}^2$ and $\Delta_{YX}^2$ as,
\begin{equation}
    \label{eqn:d}
    \Lambda = \frac{\Delta_{XY}^2 + \Delta_{YX}^2}{2}.
\end{equation}

We refer to $\Lambda$ as the Mahalanobis Distribution Alignment Score (MDAS), as it is built from Mahalanobis distances and measures how well the empirical training and test distributions are aligned in terms of their first two moments: small values indicate good alignment, while unusually large values signal distributional shift. Expanding (\ref{eqn:d}) and using the fact that $\left(n_X-1\right) / n_X \rightarrow 1$ and $\left(n_Y-1\right) / n_Y \rightarrow 1$ for large samples, we obtain the approximation
that shows $\Lambda$ is a combination of Mahalanobis distance between centroids (as in Hotelling's $T^2$ \cite{hotelling1931generalization}) and a term measuring how each sample's covariance aligns with the pooled covariance (also see Prop.~\ref{prop3}):
\begin{equation}\label{eqn:dexpanded}
\begin{aligned}
    \Lambda & \approx\left(\mu_X-\mu_Y\right)^{\top} \Sigma_p^{-1}\left(\mu_X-\mu_Y\right) \\
     & +\frac{1}{2}\left[\operatorname{tr}\left(\Sigma_p^{-1} \Sigma_X\right)+\operatorname{tr}\left(\Sigma_p^{-1} \Sigma_Y\right)\right].
\end{aligned}
\end{equation}
Since we are using a population-level average Mahalanobis distance, which is symmetric and naturally suited to the ``train vs test distributional similarity’’ idea, $\Lambda=0$ holds only in the degenerate case where every observation in both samples equals the common mean. In practice, $\Lambda \approx 0$ when the populations are very similar.

Here, $\Lambda$ is the mean of all cross-Mahalanobis squared distances, symmetrized across $X$ and $Y$, under the assumption of a shared covariance (via $\Sigma_p$). $\Lambda$ differs when the two means differ (location shift), and/or the covariance structures differ (shape/scale shift). After quantifying the distance between the training set and the test set, the question remains whether a small distance between these samples is conclusive of a good train-test split, and if so, can we infer model robustness using it? To answer this question, we devise a hypothesis test that checks whether the training set and the test set follow a similar distribution or not. In the next subsection, we formalize our strategy for hypothesis testing and provide a Monte Carlo simulation-based algorithm to implement the same. 

\subsection{Theoretical Properties of MDAS}
MDAS is a symmetric dissimilarity measure between two multivariate samples, rather than a strict metric on distributions.
\begin{proposition}[Non-negativity and symmetry]
    \begin{enumerate}
        \item $\Lambda(X, Y) \geq 0$ for all samples $X, Y$.
        \item $\Lambda(X, Y)=\Lambda(Y, X)$.
    \end{enumerate}
\end{proposition}\label{prop1}
\begin{proof}
    Each quadratic form in $\Delta_{X Y}^2, \Delta_{Y X}^2$ is non-negative because $\Sigma_p^{-1}$ is positive definite, and the average of non-negative quantities is non-negative. Symmetry follows immediately from the definition of $\Lambda$ as the average of $\Delta_{X Y}^2$ and $\Delta_{Y X}^2$.
\end{proof}
\noindent MDAS is invariant under any nonsingular affine transformation. In particular, $\Lambda$ is translation-invariant and scale/rotation invariant.

\begin{proposition}{(Affine invariance).}
    Let $A$ be any nonsingular $p \times p$ matrix and $b \in \mathbb{R}^p$. Define $$
x_{i}^{\prime}=\left\{A x_i+b\right\}, \quad y_{i}^{\prime}=\left\{A y_j+b\right\},$$
then
$$
\Lambda\left(X^{\prime}, Y^{\prime}\right)=\Lambda(X, Y).
$$
\end{proposition}\label{prop2}
\begin{proof}
    Means transform as $\mu_X^{\prime}=A \mu_X+b, \mu_Y^{\prime}=A \mu_Y+b$.
Covariances transform as $\Sigma_X^{\prime}=A \Sigma_X A^{\top}, \Sigma_Y^{\prime}=A \Sigma_Y A^{\top}$, hence $\Sigma_p^{\prime}=A \Sigma_p A^{\top}$ and $\left(\Sigma_p^{\prime}\right)^{-1}=(A^{\top})^{-1} \Sigma_p^{-1} A^{-1}$.
Each quadratic form satisfies$
\left(A x_i+b-\mu_Y^{\prime}\right)^{\top}\left(\Sigma_p^{\prime}\right)^{-1}\left(A x_i+b-\mu_Y^{\prime}\right)$ equals $\left(x_i-\mu_Y\right)^{\top} \Sigma_p^{-1}\left(x_i-\mu_Y\right),
$
and similarly for the $y_j$ 's, so all pieces of $\Lambda$ are unchanged.
\end{proof}

Now, we present Proposition 3 as the main structural property of MDAS distance.

\begin{proposition}{(Decomposition).}\label{prop3}
    Given $\Sigma_X$ and $\Sigma_Y$ denote the covariance matrices, MDAS can be decomposed into mean and covariance parts:
\begin{equation*}
\begin{aligned}
     \Lambda &=\left(\mu_X-\mu_Y\right)^{\top} \Sigma_p^{-1}\left(\mu_X-\mu_Y\right)\\
     &+\frac{1}{2}\left[\frac{n_X-1}{n_X} \operatorname{tr}\left(\Sigma_p^{-1} \Sigma_X\right)+\frac{n_Y-1}{n_Y} \operatorname{tr}\left(\Sigma_p^{-1} \Sigma_Y\right)\right],
\end{aligned}
\end{equation*}
where the first term is the Mahalanobis squared distance between sample means and the second term measures how each sample’s scatter aligns with the pooled covariance.
\end{proposition}
\begin{proof}
    For $\Delta_{X Y}^2$, write $x_i-\mu_Y=\left(x_i-\mu_X\right)+\left(\mu_X-\mu_Y\right)$. Expanding,
$$
\begin{aligned}
\Delta_{X Y}^2= & \frac{1}{n_X} \sum_i\left(x_i-\mu_X\right)^{\top} \Sigma_p^{-1}\left(x_i-\mu_X\right)\\
&+\left(\mu_X-\mu_Y\right)^{\top} \Sigma_p^{-1}\left(\mu_X-\mu_Y\right) \\
& +\frac{2}{n_X}\left(\mu_X-\mu_Y\right)^{\top} \Sigma_p^{-1} \sum_i\left(x_i-\mu_X\right)
\end{aligned}
$$
The cross term vanishes because $\sum_i\left(x_i-\mu_X\right)=0$. Using $\Sigma_X=\frac{1}{n_X-1} \sum_i\left(x_i-\mu_X\right)\left(x_i-\mu_X\right)^{\top}$, we get
$$
\frac{1}{n_X} \sum_i\left(x_i-\mu_X\right)^{\top} \Sigma_p^{-1}\left(x_i-\mu_X\right)=\frac{n_X-1}{n_X} \operatorname{tr}\left(\Sigma_p^{-1} \Sigma_X\right),
$$
so
$$
\Delta_{X Y}^2=\frac{n_X-1}{n_X} \operatorname{tr}\left(\Sigma_p^{-1} \Sigma_X\right)+\left(\mu_X-\mu_Y\right)^{\top} \Sigma_p^{-1}\left(\mu_X-\mu_Y\right) .
$$
A similar expansion holds for $\Delta_{Y X}^2$; averaging gives the formula for $\Lambda$.
\end{proof}

\noindent $\Lambda$ extends the classical Mahalanobis/Hotelling distance by adding a symmetric covariance-alignment component. From Prop.~\ref{prop3}, $\Lambda$ is at least the Mahalanobis distance between centroids. Hotelling's two-sample $T^2$ statistic is
$$
T^2=\frac{n_X n_Y}{n_X+n_Y}\left(\mu_X-\mu_Y\right)^{\top} \Sigma_p^{-1}\left(\mu_X-\mu_Y\right) .
$$
From Prop.~\ref{prop3}:
$$
\Lambda \geq\left(\mu_X-\mu_Y\right)^{\top} \Sigma_p^{-1}\left(\mu_X-\mu_Y\right)=\frac{n_X+n_Y}{n_X n_Y} T^2
$$
Thus, $\Lambda$ dominates a rescaled Hotelling distance since it combines a location term (Hotelling-type) plus a covariance mismatch term. Next, we show that $\Lambda$ is consistent as an estimator of a population-level distance between the two distributions. $\Lambda$ converges to the dimensionality plus the squared Mahalanobis distance between population means.\\

Let $\mathbf{X}=\left(\mathrm{x}_1, \mathrm{x}_2, \ldots, \mathrm{x}_{nX}\right)$ are i.i.d. samples from distribution $\mathrm{P}_{X}$ with mean $\mu_{X}$ and covariance $\Sigma_\mathrm{X}$ and $\mathbf{Y}=\left(y_1, y_2, \ldots, y_{nY}\right)$ are i.i.d. samples from distribution $P_Y$ with mean $\mu_Y$ and covariance $\Sigma_\mathrm{Y}$. Both distributions have finite fourth moments (needed for CLT applications). Define the population-level distance as:
\begin{equation*}
\begin{aligned}
    \Lambda^* &=\frac{1}{2}E_{X \sim P_X}\left[\left(\boldsymbol{X}-\boldsymbol{\mu}_{\boldsymbol{Y}}\right)^T \Sigma_p^{-1}\left(\boldsymbol{X}-\boldsymbol{\mu}_{\boldsymbol{Y}}\right)\right]\\
    &+ \frac{1}{2}E_{Y \sim P_Y}\left[\left(\boldsymbol{Y}-\boldsymbol{\mu}_{\boldsymbol{X}}\right)^T \Sigma_p^{-1}\left(\boldsymbol{Y}-\boldsymbol{\mu}_{\boldsymbol{X}}\right)\right],
\end{aligned}
\end{equation*}
where $\Sigma_{\mathrm{p}}$ is the true pooled covariance (assuming equal covariance $\Sigma_{\mathrm{X}}=\Sigma_{\mathrm{Y}}=\Sigma$).

\begin{theorem}\label{theorem1}
    Under the conditions of finite fourth moments, positive definite covariances (discussed above) and as ${n}_{\mathrm{X}}, {n}_{\mathrm{Y}} \rightarrow \infty$ with $\frac{{n}_{\mathrm{X}}}{\left({n}_{\mathrm{X}}+{n}_{\mathrm{Y}}\right)} \rightarrow \lambda \in(0,1)$, we have
    $$\Lambda \xrightarrow{p} \Lambda^*.$$
\end{theorem}
    
\begin{proof}
    By the Law of Large Numbers: $\hat{\boldsymbol{\mu}}_X \xrightarrow{p} \boldsymbol{\mu}_X$, $\hat{\boldsymbol{\mu}}_Y \xrightarrow{p} \boldsymbol{\mu}_Y$, $\hat{\Sigma}_X \xrightarrow{p} \Sigma_X$, and $\hat{\Sigma}_Y \xrightarrow{p} \Sigma_Y$.
Therefore:
$$
\hat{\Sigma}_p=\frac{\left(n_X-1\right) \hat{\Sigma}_X+\left(n_Y-1\right) \hat{\Sigma}_Y}{n_X+n_Y-2} \xrightarrow{p} \Sigma_p
$$
Combining these with Prop.~\ref{prop3} and using Slutsky’s theorem, we have
$$
\Delta_{X Y}^2 \xrightarrow{p} E_X\left[\left(\boldsymbol{X}-\boldsymbol{\mu}_Y\right)^T \Sigma_p^{-1}\left(\boldsymbol{X}-\boldsymbol{\mu}_Y\right)\right]
$$
$$
\Delta_{Y X}^2 \xrightarrow{p} E_Y\left[\left(\boldsymbol{Y}-\boldsymbol{\mu}_X\right)^T \Sigma_p^{-1}\left(\boldsymbol{Y}-\boldsymbol{\mu}_X\right)\right]
$$
By the continuous mapping theorem (average is continuous): 
\begin{equation*}
\begin{aligned}
\Lambda &=\frac{\Delta_{X Y}^2+\Delta_{Y X}^2}{2} \xrightarrow{p} \frac{1}{2}E_X\left[\left(\boldsymbol{X}-\boldsymbol{\mu}_Y\right)^T \Sigma_p^{-1}\left(\boldsymbol{X}-\boldsymbol{\mu}_Y\right)\right]\\
& +\frac{1}{2}E_Y\left[\left(\boldsymbol{Y}-\boldsymbol{\mu}_X\right)^T \Sigma_p^{-1}\left(\boldsymbol{Y}-\boldsymbol{\mu}_X\right)\right]=\Lambda^*
\end{aligned}
\end{equation*}
Under equal covariance assumption ( $\Sigma_{\mathrm{X}}=\Sigma_{\mathrm{Y}}=\Sigma$), we can simplify:
$$
\begin{gathered}
E_X\left[\left(\boldsymbol{X}-\boldsymbol{\mu}_Y\right)^T \Sigma^{-1}\left(\boldsymbol{X}-\boldsymbol{\mu}_Y\right)\right] \\
=\operatorname{tr}\left(I_p\right)+\left(\boldsymbol{\mu}_X-\boldsymbol{\mu}_Y\right)^T \Sigma^{-1}\left(\boldsymbol{\mu}_X-\boldsymbol{\mu}_Y\right) \\
=p+D^2\left(\boldsymbol{\mu}_X, \boldsymbol{\mu}_Y\right),
\end{gathered}
$$
where $p$ is the data dimensionality and $\mathrm{D}^2$ (captures genuine distributional differences) is the squared Mahalanobis distance between population means. Similar expression can be obtained for the $Y$ term and therefore, 
$\Lambda^*=p+D^2\left(\boldsymbol{\mu}_X, \boldsymbol{\mu}_Y\right)$.
\end{proof}
The consistency result (Theorem \ref{theorem1}) reveals that $\Lambda$ converges to a population-level quantity that admits an intuitive decomposition. Under the assumption of equal covariance structures, the limiting value is:
$$
\Lambda^*=p+D^2\left(\boldsymbol{\mu}_X, \boldsymbol{\mu}_Y\right).
$$
In the context of train-test split assessment, this result provides a clear decision criterion. When training and test sets are drawn from the same underlying distribution (an ideal scenario for reliable model evaluation), we expect $\Lambda \approx {p}$ for sufficiently large samples. Conversely, values of $\Lambda$ substantially exceeding $p$ indicate a distributional shift between training and test data, suggesting improper data splitting procedures or dataset drift.

\subsection{Monte Carlo Method for Hypothesis Testing}
\label{sec:hypothesis testing}

Intuitively, we formulate our null hypothesis that the training data and the test data corresponding to a good train-test split follow a similar distribution. The implicit assumption here is that when two populations are sampled from a common underlying distribution, the distance between the two populations is arbitrarily small. We use (\ref{eqn:d}) to calculate the distance between the two sets. The hypotheses can be formalized as follows.

$\boldsymbol{H_0:}$ The training data and the test data corresponding to the train-test split follow a similar distribution.
\begin{center}
\textbf{against}\\
\end{center}

$\boldsymbol{H_1:}$ The training data and the test data corresponding to the train-test split do not follow a similar distribution.

We perform a one-sided $\alpha$-level hypothesis test \cite{ruxton2010should} for our setup and calculate the value of the test statistic $\Lambda$ as $\Lambda_{obs}$, the MDAS for the given split, using (\ref{eqn:d}). To simulate the probability distribution of $\Lambda$, we run multiple simulations, i.e., repeatedly split the dataset into training and test, and calculate the MDAS for each simulation. Note that $\Lambda$ takes only positive values. Reusing the same notation, let $\Lambda$ be the random variable having the above simulated distribution. Further, we reject the null hypothesis when $\Lambda$ is greater than some constant $c > 0$. Consequently, for a given $\alpha$, we calculate the $c$ using 
\begin{equation}
    \label{eqn:rejection criterion}
    \mathbf{P}_{H_0}(\Lambda > c) \leq \alpha.
\end{equation}

\begin{figure*}
    \centering
    \includegraphics[width = \textwidth, height=12cm]{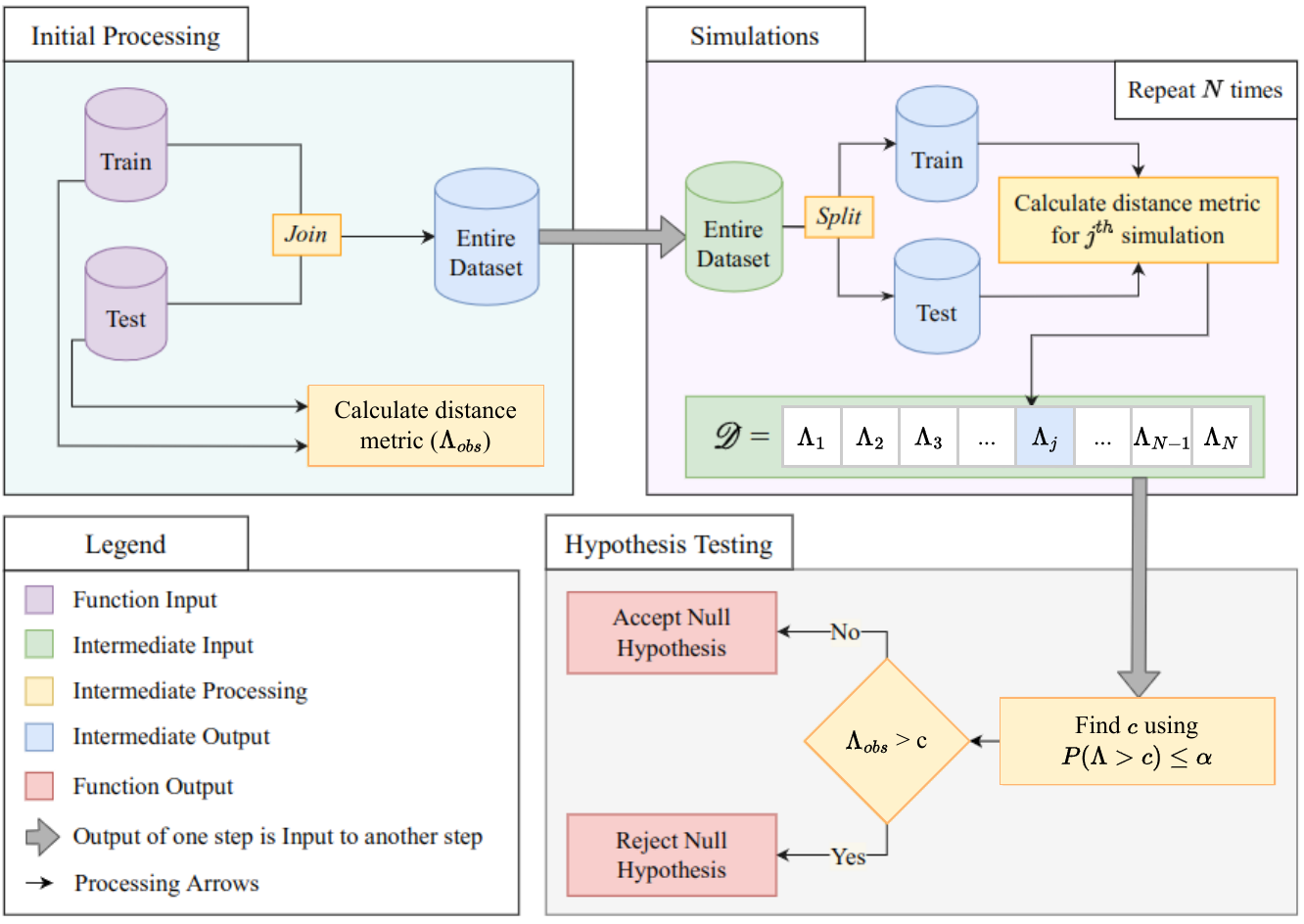}
    \caption{Monte Carlo simulation-based hypothesis testing procedure.}
    \label{fig:hypothesis testing}
\end{figure*}

Based on the rejection criterion, we will judge the quality of the random split and classify it as a good train-test split for model evaluation. We also find the p-value as follows,
\begin{equation}
    \label{eqn:pvalue}
    p = \mathbf{P}_{H_0}(\Lambda > \Lambda_{obs}).
\end{equation}
In practice, the $p$-value is estimated as
$$
\hat{p}=\frac{1+\sum_{j=1}^N \mathbf{I}\left\{\Lambda^{(j)} \geq \Lambda_{\mathrm{obs}}\right\}}{N+1},
$$
where $\Lambda^{(j)}$ are the simulated MDAS values and $\mathbf{I}$ denotes an indicator function. Fig. \ref{fig:hypothesis testing} describes the major steps of the entire process. It explains the procedure as a combination of three major steps. The first step is to calculate the test statistic $\Lambda_{obs}$ using (\ref{eqn:d}) for the input partition. Next, we obtain the complete dataset by joining the training partition and the test partition. Once we have the entire dataset, we repeatedly split (with the same split-percentage) the dataset under the random splitting paradigm and calculate the distance value for each of the random splits. After $N$ simulations, we obtain a vector of distance values for a given dataset. If $N$ is large enough, we can assume that this vector simulates the probability distribution of the test statistic $\Lambda$. Finally we reject the null hypothesis if $\Lambda_{obs} > c$. Algorithm (\ref{alg:simulate}) describes the entire process. 

\begin{algorithm}
 \caption{Monte Carlo Simulation-based Hypothesis Test}
 \begin{algorithmic}[1]\label{alg:simulate}
 \renewcommand{\algorithmicrequire}{\textbf{Require:}}
 \REQUIRE $X$, $Y$, $\alpha$, $N$
  \STATE $\Lambda_{obs} =$ calculate$\_$distance($X$, $Y$)
  \STATE Initialize $\mathcal{D}[1, \dots, N]$ to store the MDAS values
  \STATE Join $X$ and $Y$ sets to obtain the entire dataset ($df$)
  \STATE $s = $ calculate$\_$split$\_$percentage($X$, $Y$) 
  \FOR{$j = 1$ \TO $N$}
        \STATE $(X,\ Y) = $ random$\_$split($df$, $s$)
        \STATE $\Lambda = $ calculate$\_$distance($X$, $Y$)
        \STATE $\mathcal{D}[j] = \Lambda$
        \ENDFOR
   \STATE $p =$ calculate$\_$pvalue($\mathcal{D}$, $\Lambda_{obs}$) 
    \IF{$p > \alpha$}
    \STATE Accept Null Hypothesis 
    \ELSE
    \STATE Reject Null Hypothesis 
    \ENDIF
 \end{algorithmic} 
 \end{algorithm}

\subsection{Association of model performance}
\label{subsec:naic}
Generally, there are several attributes (or features) in real-life datasets. Choosing a subset of these attributes to establish a definite model relation a priori is unexpected and uncommon. In this light, we presented the above technique, which works on the entire dataset without assuming any learning objective or model relation or fitting any regression/classification models. However, if the model relation is provided, the proposed method can associate model performance with the proposed MDAS through a simulation plot. The Akaike Information Criterion \cite{sakamoto1986akaike, burnham1998practical} is used as the performance metric as,  
\begin{equation}\label{eqn:AIC}
    {AIC} = -2 \log(\mathcal{L}(\hat{\theta} \mid  x)) + 2K
\end{equation}
where $\mathcal{L}$ is the maximum value of the likelihood function for the model, and $K$ is the number of parameters in the model. In ordinary least squares regression, the residual sum of squares (RSS) \cite{draper1998applied} is calculated as,
\begin{equation}
    \label{eqn:RSS}
    {RSS} = \sum_{i = 1}^{n} (y_i - \hat{y}_i)^2
\end{equation}
where $y_i$ is the $i^{th}$ observed value of the response variable, and $\hat{y}_i$ is the $i^{th}$ predicted value of the response variable. Thus, in the case of ordinary least squares regression \cite{burnham2011aic}, 
\begin{equation}
    \label{eqn:max log likelihood}
    \log (\mathcal{L}) = - \Big(\frac{n}{2}\Big) \log \Big(\frac{{RSS}}{n}\Big),
\end{equation}
which gives 
\begin{equation}
    \label{eqn:AIC RSS}
    {AIC} =  n \log \Big(\frac{{RSS}}{n}\Big) + 2K.
\end{equation}
Since AIC is dependent on the sample size $n$, we will use the normalized form of the metric to make it invariant to sample size \cite{cohen2021normalized}. The normalized AIC is obtained by dividing the AIC score by the sample size and can be expressed as,
\begin{equation}
    \label{eqn:normalized AIC}
    {AIC_N} = \log \Big(\frac{{RSS}}{n}\Big) + \frac{2K}{n}.
\end{equation}
For visualizing different simulations (data splits), we repeatedly split the dataset, train the model using the given model information, calculate MDAS between the training and test sets, and measure the model performances for both sets using (\ref{eqn:normalized AIC}). In the next section, we provide examples through real-life regression datasets when the model information is provided and when it is not.

\section{Experiments and Results}
\label{sec:experiments}

We consider regression analysis to present the findings computationally. The datasets used for the experiment are discussed below. For all the datasets, we run the Algorithm (\ref{alg:simulate}) and obtain a conclusion regarding the quality of the random split using R.  

\textbf{Abalone}: This dataset is taken from the UCI Machine Learning Repository \cite{Dua:2019}.  This data came from an original study conducted by WJ Nash \cite{nash19947he}. There are $9$ variables, out of which one is an ordered factor, one is an integer, and the rest are continuous variables. The predominant purpose of the dataset is to predict the age of abalone, characterized by the variable Rings, from their physical measurements. We consider a regression model with \textit{Rings}  as the response variable and independent variables as \textit{LongestShell}, \textit{Diameter}, and  \textit{Height}. Here, \textit{LongestShell} denotes the maximum length of the shell of abalone, the \textit{Diameter} is the length perpendicular to the longest shell, and \textit{Height} is the height of abalone. The information has been tabulated in Table \ref{tab:datasets}.

\begin{table*}[htbp]
    \centering
    \caption{Information about the used datasets.}
    \begin{tabular}{llr}
\hline
Attribute & Abalone dataset & Diamonds dataset \\
\hline
    No. of Rows (Sample Size)  & 4,177  & 53,940\\ 
        No. of Columns (Variables)  & 9  & 10\\ 
        Model Relation (Regression) & Rings $\sim$ LongestShell + Diameter + Height  & price $\sim$ volume + depth\\ 
\hline
\end{tabular}
    \label{tab:datasets}
\end{table*}

\textbf{Diamonds}: It is available in the ggplot2 library in R. There are a total of $10$ variables, out of which three are ordered factors, one is an integer, and the remaining six are numeric. These variables measure the various characteristics of $53,940$ round-cut diamonds. We define the regression model with \textit{price} as the response variable and independent variables as \textit{depth}, \textit{x:y:z}. Here, \textit{price} denotes the price of the diamond in US dollars, \textit{depth} denotes the total depth percentage, and \textit{x}, \textit{y}, and \textit{z} denote the length, width, and height in millimeters, respectively. The product, \textit{x:y:z}, of the three dimensions \textit{x}, \textit{y}, and \textit{z} can be interpreted as the volume of the diamond. The above relation precisely conveys that the price of the diamond is a linear combination of the depth percentage of the diamond and its volume. The above information has been tabulated in Table \ref{tab:datasets}.

\subsection{Models and Evaluations}
\label{sec:Models}
\begin{figure*}[!ht]
    \centering
    \begin{subfigure}[b]{0.85\textwidth}
        \centering
        \includegraphics[width = \textwidth, height=5cm]{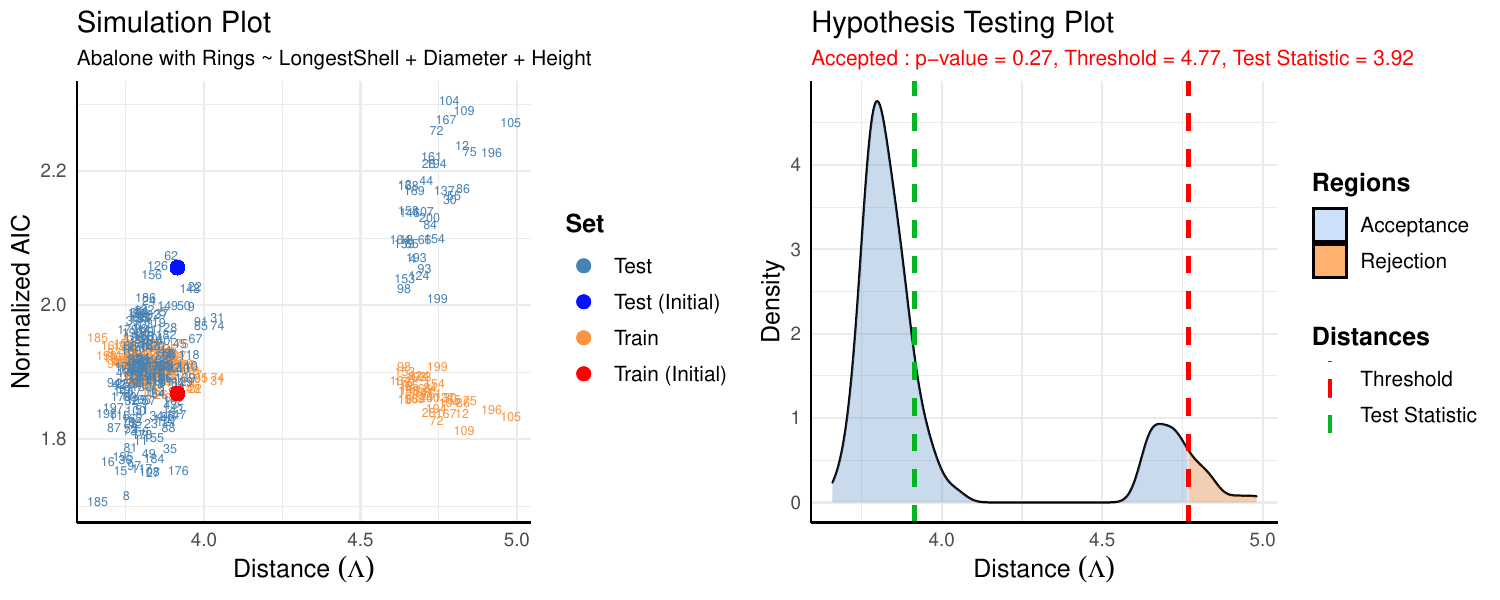}
        \caption{Abalone Dataset for seed $= 3$ (null hypothesis accepted).}
        \label{fig:abalone acc}    
    \end{subfigure}
    
    \begin{subfigure}[b]{0.85\textwidth}
        \centering
        \includegraphics[width = \textwidth, height=5cm]{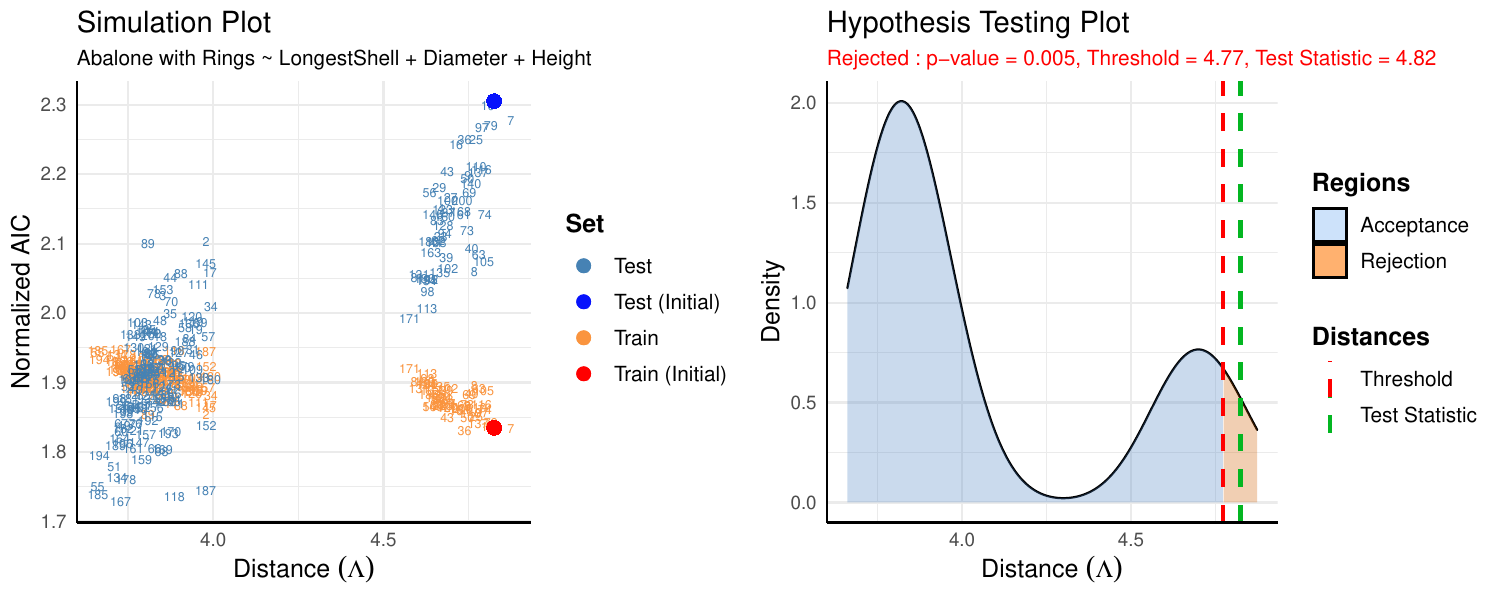}
        \caption{Abalone Dataset for seed $= 20$ (null hypothesis rejected).}
        \label{fig:abalone rej}    
    \end{subfigure}
    
    \caption{Simulations for Abalone Dataset. We regress the Rings of abalone using the longest shell, diameter, and the height of the abalone.}
\end{figure*}

We present four examples, two for each dataset. Fig. \ref{fig:abalone acc} shows a random split with seed as $3$. Using Algorithm (\ref{alg:simulate}) and comparing the given split with approximately all possible splits, the proposed method accepts the null hypothesis to conclude that the training set distribution and the test set distribution are similar. Accepting the null hypothesis indicates that the model performance measured corresponding to the generated split is reliable. Fig. \ref{fig:abalone rej} shows another simulation for the abalone dataset with seed as $20$. We observe that the null hypothesis is rejected as the split lies in the right-most clusters in the simulation plot of Fig. \ref{fig:abalone rej}. According to our analysis, this split is not a good split to assess model performance as it is a corner case, and a poor model performance on such a split doesn't signify a poor model. Although, a good model performance on such a split can ensure model robustness. Table \ref{tab:stats} summarizes the two simulations.

\setlength{\tabcolsep}{10pt}
\begin{table*}[!ht]
    \caption{Conclusion table for simulations on both datasets.}
    \centering
    \label{tab:stats}
    \begin{footnotesize}
    \begin{tabular}{lllll}
        \toprule
        \multirow{2}{*}{Attribute} & \multicolumn{2}{c}{Abalone dataset} & \multicolumn{2}{c}{Diamonds dataset}  \\ 
        & Run 1 & Run 2 & Run 1 & Run 2 \\
        \midrule
        R seed & 3 & 20 & 2 & 1\\
        MDAS ($\Lambda$) & 3.912  & 4.825 & 2.911   & 3.331 \\ 
        Limiting Threshold (c) & 4.768  & 4.772 & 3.324  & 3.319\\ 
        p-value & 0.27 & 0.005 & 0.845 & 0.025 \\
        Model Performance for Initial Train Split & 1.868  & 1.835 & 14.941  & 14.707\\ 
        Model Performance for Initial Test Split & 2.056 & 2.305 & 14.704 & 15.444\\ 
        Split Conclusion & Accepted  & Rejected & Accepted  & Rejected\\ 
        \bottomrule
    \end{tabular}
    \end{footnotesize}
\end{table*}

\begin{figure*}[!ht]
    \centering
    \begin{subfigure}[b]{0.90\textwidth}
        \centering
        \includegraphics[width = \textwidth, height=5cm]{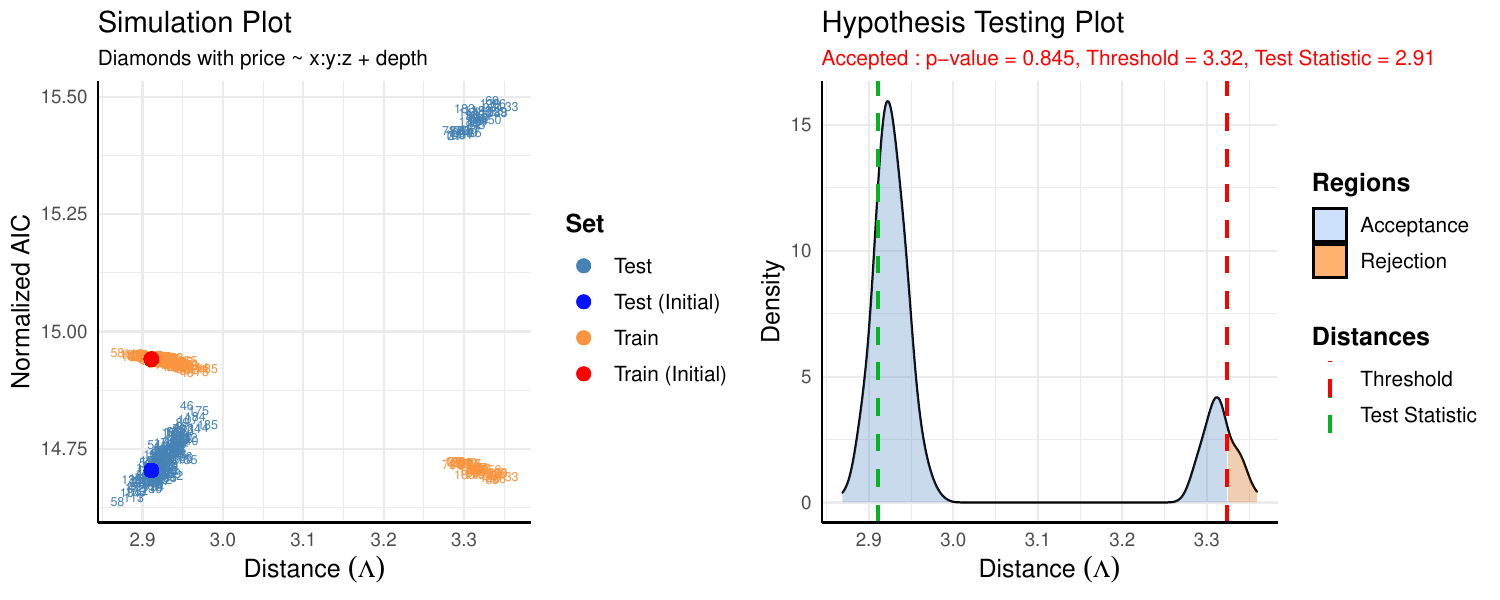}
        \caption{Diamonds Dataset for seed $= 2$ (null hypothesis accepted).}
        \label{fig:diamonds acc}
    \end{subfigure}
    
    \begin{subfigure}[b]{0.90\textwidth}
        \centering
        \includegraphics[width = \textwidth, height=5cm]{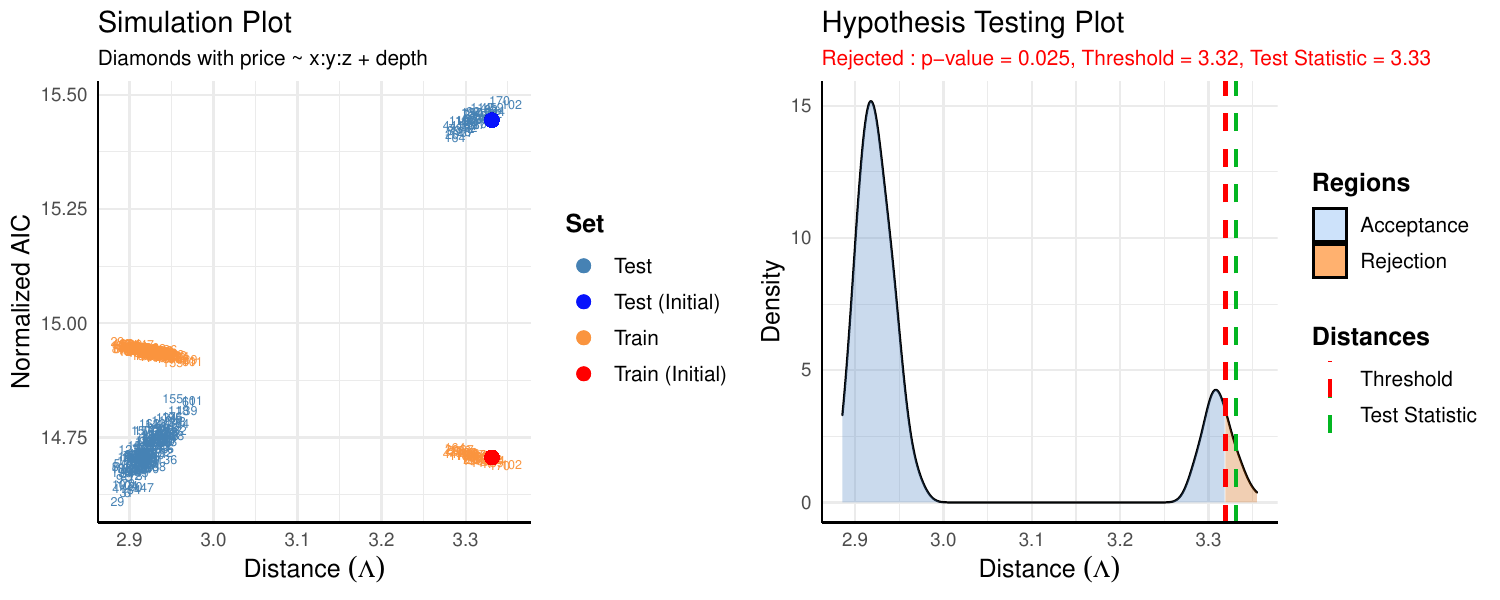}
        \caption{Diamonds Dataset for seed $= 1$ (null hypothesis rejected).}
        \label{fig:diamonds rej}
    \end{subfigure}
    
    \caption{Simulations for Diamonds Dataset. The price of the diamond is regressed using the volume (x:y:z) and the depth of the diamond.}
\end{figure*}

A similar experimental analysis for the Diamonds dataset is held. A random split with seed $2$ results in the null hypothesis being accepted. The simulation is visualized if Fig. \ref{fig:diamonds acc}. The random split lies in the left region of the simulation plot of Fig. \ref{fig:diamonds acc}, indicating a small distance between the training set and the test set. We conclude that this split can be used for measuring model performance. On the other hand, a random split with seed $1$ ends up being rejected since the distance between the training set and the test set is large. This split is not ideal for measuring model performance since it is a corner case and can potentially underestimate model performances even for a good model. The results for the two simulations have been collected in Table \ref{tab:stats}

\subsection{Comparison with Existing Data Splitting Methods}

\begin{figure*}[!ht]
    \centering
    \begin{subfigure}[b]{0.85\textwidth}
        \centering
        \includegraphics[width = \textwidth, height=5cm]{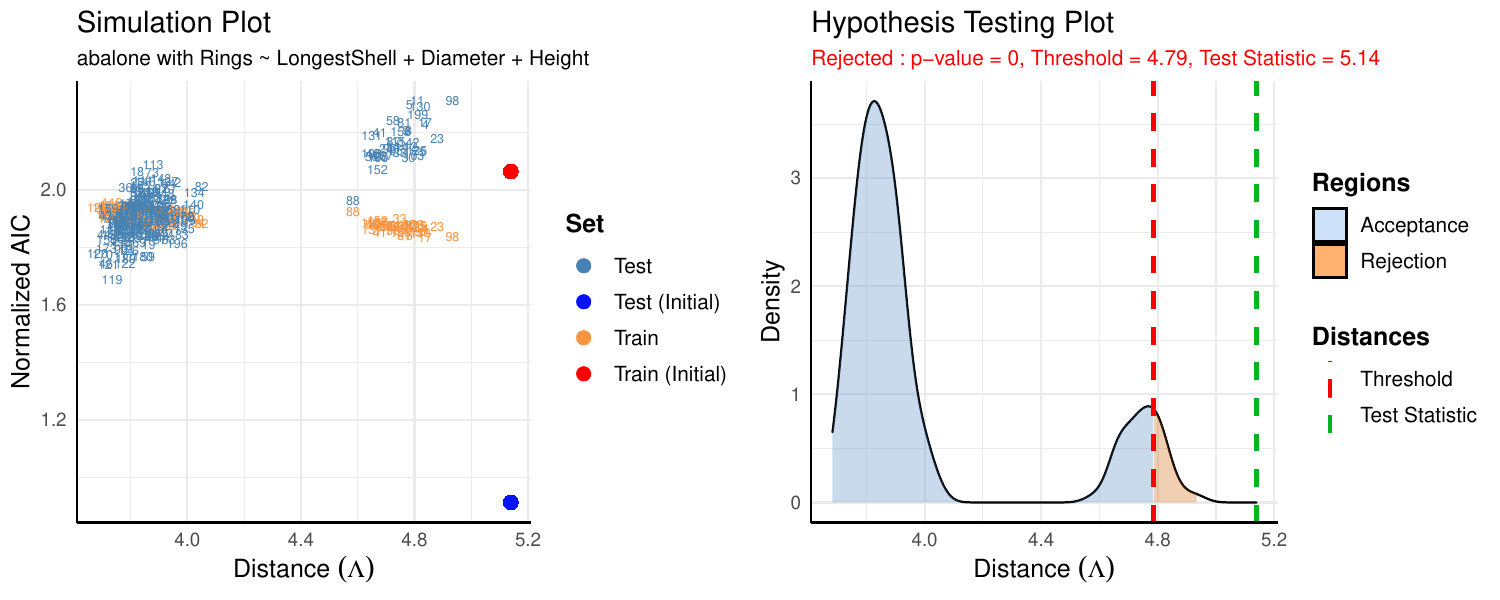}
        \caption{Abalone Dataset with CADEX Splitting (null hypothesis rejected).}
        \label{fig:cadex}
    \end{subfigure}
    
    \begin{subfigure}[b]{0.85\textwidth}
        \centering
        \includegraphics[width = \textwidth, height=5cm]{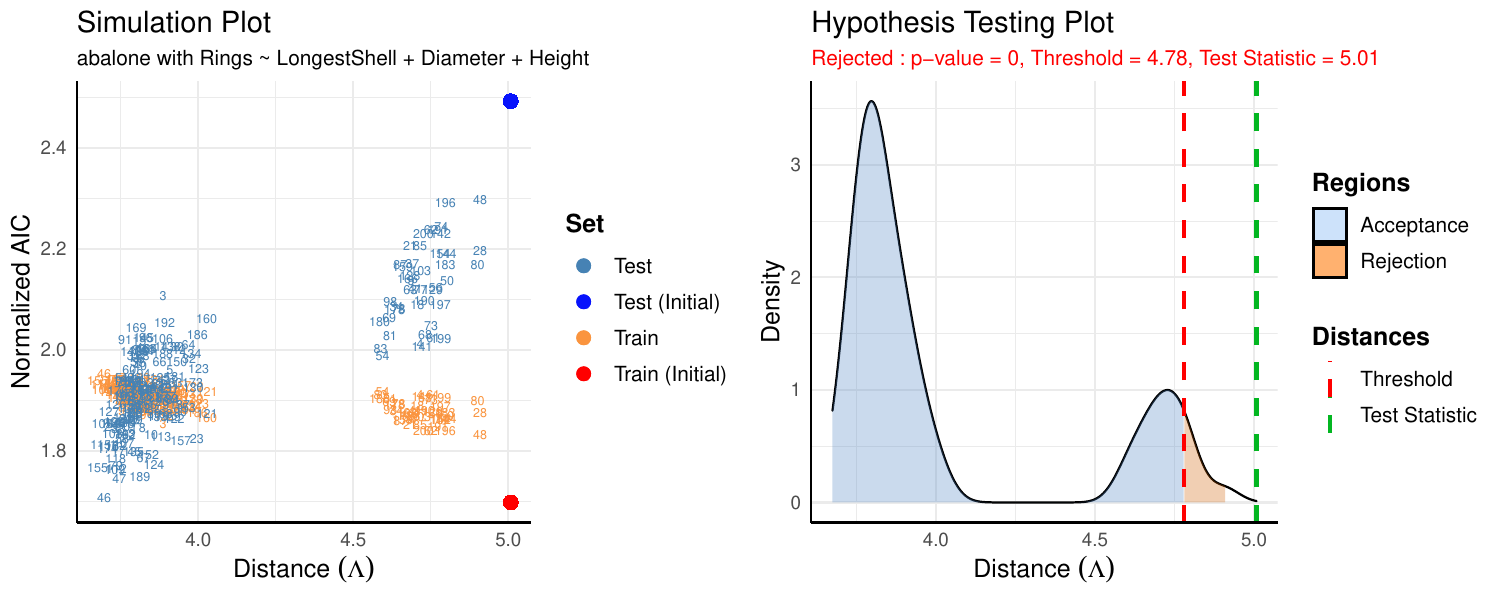}
        \caption{Abalone Dataset with DUPLEX Splitting (null hypothesis rejected).}
        \label{fig:duplex}
    \end{subfigure}
    
    \begin{subfigure}[b]{0.85\textwidth}
        \centering
        \includegraphics[width = \textwidth, height=5cm]{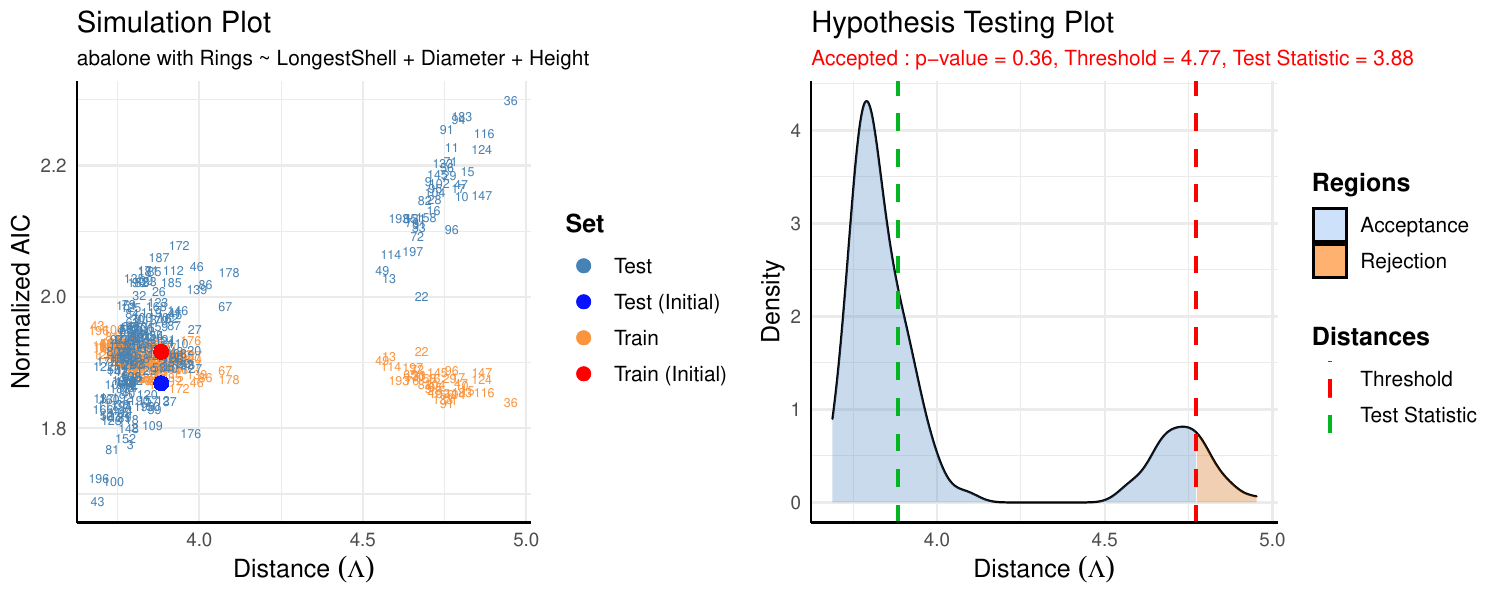}
        \caption{Abalone Dataset with SPlit splitting (null hypothesis accepted).}
        \label{fig:split}
    \end{subfigure}
    \caption{Comparing different splitting techniques using our hypothesis testing method.}
\end{figure*}

We also compare existing data splitting strategies like SPlit \cite{joseph2021split}, CADEX \cite{kennard1969computer}, and DUPLEX \cite{snee1977validation} on the Abalone dataset. We compare the splits produced by these methods among the other possible splits. We visualize the presence of the initial split and conclude its appropriateness based on the hypothesis testing method discussed in Section \ref{sec:hypothesis testing}. In comparing, we observe that splits produced by the CADEX (Fig. \ref{fig:cadex}) and DUPLEX (Fig. \ref{fig:duplex}) subsampling methods are rejected by our hypothesis testing method. This points out that the split obtained through these methods is not ideal for measuring model performance. On the other hand, the SPlit method developed by Joseph and Vakayil \cite{joseph2021split}, does produce an acceptable split (Fig. \ref{fig:split}). However, concluding model robustness from SPlit's split is not recommended; it may overestimate the model performance owing to the equitable representation of the entire data in the test set. 

\section{Conclusions and Future Work}
\label{sec:conclusions}

Random splitting is the most common method used for data splitting in machine learning tasks. The proposed method includes a data-driven distance, MDAS, based on the Mahalanobis squared distance. We simulate the distribution of the MDAS by repeatedly splitting the data in a random manner and calculating the corresponding distance. We then impose an $\alpha$-level one-sided hypothesis test with the null hypothesis stating that the training set and the test set of a train-test split follow a similar distribution. The proposed method diagnoses a given split among all possible splits for that dataset. Further, we compare various existing data splitting techniques using the proposed method and discuss whether the splits produced by them are good or not for measuring reliable model performance. The ability of our method to gauge the "goodness" of any given split among all other possible splits is one of a kind. We provide a diagnostic approach to assess the quality of a suitable split based on the type of problem at hand. Our method can also be used to judge train-validation splits by changing the initial split input to the algorithm. There is scope for research to extend the proposed method to consider ordinal and nominal variables by using a generalized form of Mahalanobis squared distance \cite{de2005generalized}.

We have applied our method to several regression datasets using different choices of model relations and found that it accurately diagnoses the input splits. The use of Monte Carlo simulations in the hypothesis test allows the method invariant to the dataset. Due to its dynamic nature, the proposed method is valid not only for random splits but also for any given adversarial split. Finally, the proposed method assesses the quality of a train-test data split without considering any model relation. However, if the model is specified, it can also compare the relative performances of the model in training and test data concerning all possible splits. 

The proposed Mahalanobis Distribution Alignment Score is sensitive to differences in the mean vectors of the training and test sets, and will also flag situations in which observations from one set behave as outliers relative to the other. However, it relies on the assumption that the two samples share a common covariance structure, and a small value of $\Lambda$ should be interpreted as indicating similarity in location and covariance rather than full distributional equivalence. Moreover, in high-dimensional settings, the estimation of the pooled covariance matrix can be unstable, which may affect the reliability of the resulting score.

\section*{Data and Code Availability Statement}
Data sets analysed in this study are taken from UCI ML Repository \cite{Dua:2019} and \texttt{ggplot2} library in R. An R package has been developed to implement the proposed methodology easily for users and is available at \url{https://github.com/eklavyaj/RandomSplitDiagnostics}.

\bibliographystyle{abbrvnat}
\bibliography{References}

@article{yu2025generating,
  title={Generating samples for covariance to update prototype in few-shot class-incremental learning},
  author={Yu, Hong and Luo, Qiwei and Wang, Ye and Wang, Guoyin},
  journal={Applied Intelligence},
  volume={55},
  number={18},
  pages={1126},
  year={2025},
  publisher={Springer}
}

@article{kahloot2021algorithmic,
  title={Algorithmic splitting: A method for dataset preparation},
  author={Kahloot, Khalid M and Ekler, Peter},
  journal={IEEE access},
  volume={9},
  pages={125229--125237},
  year={2021},
  publisher={IEEE}
}

@inproceedings{sogaard2020we,
  title={We need to talk about random splits},
  author={S{\o}gaard, Anders and Ebert, Sebastian and Bastings, Jasmijn and Filippova, Katja},
  booktitle={Proceedings of the 16th conference of the European chapter of the association for computational linguistics: main volume},
  pages={1823--1832},
  year={2021}
}

@article{zhang2011image,
  title={Image segmentation using PSO and PCM with Mahalanobis distance},
  author={Zhang, Yong and Huang, Dan and Ji, Min and Xie, Fuding},
  journal={Expert systems with applications},
  volume={38},
  number={7},
  pages={9036--9040},
  year={2011},
  publisher={Elsevier}
}

@article{bzdok2018krzywinski,
  title={Statistics versus machine learning},
  author={Bzdok, Danilo and Altman, Naomi and Krzywinski, M},
  journal={Nature Methods},
  volume={15},
  number={04},
  pages={233-234},
  year={2018}
}

@book{zellner2004statistics,
  title={Statistics, econometrics and forecasting},
  author={Zellner, Arnold},
  year={2004},
  publisher={Cambridge University Press}
}

@article{liu2025thompson,
  title={Thompson sampling for zero-inflated count outcomes with an application to the Drink Less mobile health study},
  author={Liu, Xueqing and Deliu, Nina and Chakraborty, Tanujit and Bell, Lauren and Chakraborty, Bibhas},
  journal={The Annals of Applied Statistics},
  volume={19},
  number={2},
  pages={1403--1425},
  year={2025},
  publisher={Institute of Mathematical Statistics}
}

@article{altalhan2025imbalanced,
  title={Imbalanced data problem in machine learning: A review},
  author={Altalhan, Manahel and Algarni, Abdulmohsen and Alouane, Monia Turki-Hadj},
  journal={IEEE Access},
  year={2025},
  publisher={IEEE}
}

@article{li2025machine,
  title={Machine unlearning: Taxonomy, metrics, applications, challenges, and prospects},
  author={Li, Na and Zhou, Chunyi and Gao, Yansong and Chen, Hui and Zhang, Zhi and Kuang, Boyu and Fu, Anmin},
  journal={IEEE Transactions on Neural Networks and Learning Systems},
  year={2025},
  publisher={IEEE}
}

@article{zha2025data,
  title={Data-centric artificial intelligence: A survey},
  author={Zha, Daochen and Bhat, Zaid Pervaiz and Lai, Kwei-Herng and Yang, Fan and Jiang, Zhimeng and Zhong, Shaochen and Hu, Xia},
  journal={ACM Computing Surveys},
  volume={57},
  number={5},
  pages={1--42},
  year={2025},
  publisher={ACM New York, NY}
}

@article{vamathevan2019applications,
  title={Applications of machine learning in drug discovery and development},
  author={Vamathevan, Jessica and Clark, Dominic and Czodrowski, Paul and Dunham, Ian and Ferran, Edgardo and Lee, George and Li, Bin and Madabhushi, Anant and Shah, Parantu and Spitzer, Michaela and others},
  journal={Nature reviews Drug discovery},
  volume={18},
  number={6},
  pages={463--477},
  year={2019},
  publisher={Nature Publishing Group UK London}
}

@article{varoquaux2023evaluating,
  title={Evaluating machine learning models and their diagnostic value},
  author={Varoquaux, Ga{\"e}l and Colliot, Olivier},
  journal={Machine learning for brain disorders},
  pages={601--630},
  year={2023},
  publisher={Springer}
}

@article{babaei2025impact,
  title={The impact of data splitting methods on machine learning models: A case study for predicting concrete workability},
  author={Babaei, Hossein and Zamani, Mohammad and Mohammadi, Soheil},
  journal={Machine Learning for Computational Science and Engineering},
  volume={1},
  number={1},
  pages={21},
  year={2025},
  publisher={Springer}
}

@inproceedings{reitermanova2010data,
  title={Data splitting},
  author={Reitermanova, Zuzana and others},
  booktitle={WDS},
  volume={10},
  pages={31--36},
  year={2010},
  organization={Matfyzpress Prague}
}

@article{l1991implementing,
  title={Implementing a random number package with splitting facilities},
  author={L'Ecuyer, Pierre and Cote, Serge},
  journal={ACM Transactions on Mathematical Software (TOMS)},
  volume={17},
  number={1},
  pages={98--111},
  year={1991},
  publisher={ACM New York, NY, USA}
}

@article{ishwaran2015effect,
  title={The effect of splitting on random forests},
  author={Ishwaran, Hemant},
  journal={Machine learning},
  volume={99},
  number={1},
  pages={75--118},
  year={2015},
  publisher={Springer}
}

@article{sadhukhan2024footprints,
  title={Footprints of Data in a Classifier: Understanding the Privacy Risks and Solution Strategies},
  author={Sadhukhan, Payel and Chakraborty, Tanujit},
  journal={arXiv preprint arXiv:2407.02268},
  year={2024}
}

@article{hotelling1931generalization,
author = {Harold Hotelling},
title = {{The Generalization of Student's Ratio}},
volume = {2},
journal = {The Annals of Mathematical Statistics},
number = {3},
publisher = {Institute of Mathematical Statistics},
pages = {360 -- 378},
year = {1931},
doi = {10.1214/aoms/1177732979},
URL = {https://doi.org/10.1214/aoms/1177732979}
}

@book{friedman2015fundamentals,
  title={Fundamentals of clinical trials},
  author={Friedman, Lawrence M and Furberg, Curt D and DeMets, David L and Reboussin, David M and Granger, Christopher B},
  year={2015},
  publisher={Springer}
}

@article{joseph2021split,
  title={Split: An optimal method for data splitting},
  author={Joseph, V Roshan and Vakayil, Akhil},
  journal={Technometrics},
  pages={1-11},
  year={2021},
  publisher={Taylor \& Francis}
}

@book{hastie2009elements,
  title={The elements of statistical learning: data mining, inference, and prediction},
  author={Hastie, Trevor and Tibshirani, Robert and Friedman, Jerome H and Friedman, Jerome H},
  volume={2},
  year={2009},
  publisher={Springer}
}

@article{cohen2021normalized,
  title={Normalized Information Criteria and Model Selection in the Presence of Missing Data},
  author={Cohen, Nitzan and Berchenko, Yakir},
  journal={Mathematics},
  volume={9},
  number={19},
  pages={2474},
  year={2021},
  publisher={Multidisciplinary Digital Publishing Institute}
}

@article{mclachlan1999mahalanobis,
  title={Mahalanobis distance},
  author={McLachlan, Goeffrey J},
  journal={Resonance},
  volume={4},
  number={6},
  pages={20--26},
  year={1999},
  publisher={Springer India, in co-publication with Indian Academy of Sciences}
}

@article{xiang2008learning,
  title={Learning a Mahalanobis distance metric for data clustering and classification},
  author={Xiang, Shiming and Nie, Feiping and Zhang, Changshui},
  journal={Pattern recognition},
  volume={41},
  number={12},
  pages={3600--3612},
  year={2008},
  publisher={Elsevier}
}

@article{krishna1999genetic,
  title={Genetic K-means algorithm},
  author={Krishna, K and Murty, M Narasimha},
  journal={IEEE Transactions on Systems, Man, and Cybernetics, Part B (Cybernetics)},
  volume={29},
  number={3},
  pages={433--439},
  year={1999},
  publisher={IEEE}
}

@article{peterson2009k,
  title={K-nearest neighbor},
  author={Peterson, Leif E},
  journal={Scholarpedia},
  volume={4},
  number={2},
  pages={1883},
  year={2009}
}

@article{doi:10.1080/00401706.2014.902774,
    author = {Pedro Galeano and Esdras Joseph and Rosa E. Lillo},
    title = {The Mahalanobis Distance for Functional Data With Applications to Classification},
    journal = {Technometrics},
    volume = {57},
    number = {2},
    pages = {281-291},
    year  = {2015},
    publisher = {Taylor & Francis},
    doi = {10.1080/00401706.2014.902774},
    
    URL = { 
            https://doi.org/10.1080/00401706.2014.902774
        
    },
    eprint = { 
            https://doi.org/10.1080/00401706.2014.902774
        
    }

}

@article{geun2000multivariate,
  title={Multivariate outliers and decompositions of Mahalanobis distance},
  author={Geun Kim, Myung},
  journal={Communications in statistics-theory and methods},
  volume={29},
  number={7},
  pages={1511--1526},
  year={2000},
  publisher={Taylor \& Francis}
}

@article{brereton2016re,
  title={Re-evaluating the role of the Mahalanobis distance measure},
  author={Brereton, Richard G and Lloyd, Gavin R},
  journal={Journal of Chemometrics},
  volume={30},
  number={4},
  pages={134--143},
  year={2016},
  publisher={Wiley Online Library}
}

@article{stone1974cross,
  title={Cross-validatory choice and assessment of statistical predictions},
  author={Stone, Mervyn},
  journal={Journal of the royal statistical society: Series B (Methodological)},
  volume={36},
  number={2},
  pages={111--133},
  year={1974},
  publisher={Wiley Online Library}
}

@article{theiler1996constrained,
  title={Constrained-realization Monte-Carlo method for hypothesis testing},
  author={Theiler, James and Prichard, Dean},
  journal={Physica D: Nonlinear Phenomena},
  volume={94},
  number={4},
  pages={221--235},
  year={1996},
  publisher={Elsevier}
}

@article{sakamoto1986akaike,
  title={Akaike information criterion statistics},
  author={Sakamoto, Yosiyuki and Ishiguro, Makio and Kitagawa, Genshiro},
  journal={Dordrecht, The Netherlands: D. Reidel},
  volume={81},
  number={10.5555},
  pages={26853},
  year={1986},
  publisher={Taylor \& Francis}
}

@book{seber2009multivariate,
  title={Multivariate observations},
  author={Seber, George AF},
  year={2009},
  publisher={John Wiley \& Sons}
}

@incollection{burnham1998practical,
  title={Practical use of the information-theoretic approach},
  author={Burnham, Kenneth P and Anderson, David R},
  booktitle={Model selection and inference},
  pages={75--117},
  year={1998},
  publisher={Springer}
}

@article{burnham2011aic,
  title={AIC model selection and multimodel inference in behavioral ecology: some background, observations, and comparisons},
  author={Burnham, Kenneth P and Anderson, David R and Huyvaert, Kathryn P},
  journal={Behavioral ecology and sociobiology},
  volume={65},
  number={1},
  pages={23--35},
  year={2011},
  publisher={Springer}
}

@book{draper1998applied,
  title={Applied regression analysis},
  author={Draper, Norman R and Smith, Harry},
  volume={326},
  year={1998},
  publisher={John Wiley \& Sons}
}

@article{ruxton2010should,
  title={When should we use one-tailed hypothesis testing?},
  author={Ruxton, Graeme D and Neuh{\"a}user, Markus},
  journal={Methods in Ecology and Evolution},
  volume={1},
  number={2},
  pages={114--117},
  year={2010}
}

@misc{Dua:2019 ,
author = "Dua, Dheeru and Graff, Casey",
year = "2017",
title = "{UCI} Machine Learning Repository",
url = "http://archive.ics.uci.edu/ml",
institution = "University of California, Irvine, School of Information and Computer Sciences" }

@article{nash19947he,
  title={7he population biology of abalone (\_Haliotis\_ species) in Tasmania. I},
  author={Nash, WJ},
  journal={Blacklip abalone (\_H. rubra\_) from the North Coast and Islands of Bass Strait},
  year={1994}
}

@article{osborne2004power,
  title={The power of outliers (and why researchers should always check for them)},
  author={Osborne, Jason W and Overbay, Amy},
  journal={Practical Assessment, Research, and Evaluation},
  volume={9},
  number={1},
  pages={6},
  year={2004}
}

@misc{birba2020comparative,
  title={A Comparative study of data splitting algorithms for machine learning model selection},
  author={Birba, Delwende Eliane},
  year={2020}
}

@inproceedings{kohavi1995study,
  title={A study of cross-validation and bootstrap for accuracy estimation and model selection},
  author={Kohavi, Ron and others},
  booktitle={Ijcai},
  volume={14},
  number={2},
  pages={1137--1145},
  year={1995},
  organization={Montreal, Canada}
}

@article{kennard1969computer,
  title={Computer aided design of experiments},
  author={Kennard, Ronald W and Stone, Larry A},
  journal={Technometrics},
  volume={11},
  number={1},
  pages={137--148},
  year={1969},
  publisher={Taylor \& Francis}
}

@article{GALVAO2005736,
title = {A method for calibration and validation subset partitioning},
journal = {Talanta},
volume = {67},
number = {4},
pages = {736-740},
year = {2005},
issn = {0039-9140},
doi = {https://doi.org/10.1016/j.talanta.2005.03.025},
url = {https://www.sciencedirect.com/science/article/pii/S003991400500192X},
author = {Roberto Kawakami Harrop Galvão and Mário César Ugulino Araujo and Gledson Emídio José and Marcio José Coelho Pontes and Edvan Cirino Silva and Teresa Cristina Bezerra Saldanha}
}

@article{xu2018splitting,
  title={On splitting training and validation set: a comparative study of cross-validation, bootstrap and systematic sampling for estimating the generalization performance of supervised learning},
  author={Xu, Yun and Goodacre, Royston},
  journal={Journal of analysis and testing},
  volume={2},
  number={3},
  pages={249--262},
  year={2018},
  publisher={Springer}
}

@article{refaeilzadeh2009cross,
  title={Cross-validation.},
  author={Refaeilzadeh, Payam and Tang, Lei and Liu, Huan},
  journal={Encyclopedia of database systems},
  volume={5},
  pages={532--538},
  year={2009},
  publisher={Springer}
}

@article{snee1977validation,
  title={Validation of regression models: methods and examples},
  author={Snee, Ronald D},
  journal={Technometrics},
  volume={19},
  number={4},
  pages={415--428},
  year={1977},
  publisher={Taylor \& Francis}
}

@article{de2005generalized,
  title={A generalized Mahalanobis distance for mixed data},
  author={De Leon, AR and Carriere, KC},
  journal={Journal of Multivariate Analysis},
  volume={92},
  number={1},
  pages={174--185},
  year={2005},
  publisher={Elsevier}
}

@inproceedings{mahalanobis1936generalized,
  title={On the generalized distance in statistics},
  author={Mahalanobis, Prasanta Chandra},
  year={1936},
  organization={National Institute of Science of India}
}

@article{picard1990data,
  title={Data splitting},
  author={Picard, Richard R and Berk, Kenneth N},
  journal={The American Statistician},
  volume={44},
  number={2},
  pages={140--147},
  year={1990},
  publisher={Taylor \& Francis}
}
\end{document}